\documentclass[longauth]{aa}

\usepackage{epsfig}
\usepackage{graphicx}
\usepackage{amssymb}
\usepackage{rotating}

\usepackage{natbib}
\bibpunct{(}{)}{;}{a}{}{,} 

\begin{document}

\newlength\imageheight
\setlength{\imageheight}{\textheight}
\addtolength{\imageheight}{-0.8cm}

\title{Detection of VHE gamma-ray emission from the distant blazar 1ES\,1101-232 with H.E.S.S.
and broadband characterisation}

\author{F. Aharonian\inst{1}
 \and A.G.~Akhperjanian \inst{2}
 \and A.R.~Bazer-Bachi \inst{3}
 \and M.~Beilicke \inst{4}
 \and W.~Benbow \inst{1}
 \and D.~Berge \inst{1} \thanks{now at CERN, Geneva, Switzerland}
 \and K.~Bernl\"ohr \inst{1,5}
 \and C.~Boisson \inst{6}
 \and O.~Bolz \inst{1}
 \and V.~Borrel \inst{3}
% ...
}

\institute{
Max-Planck-Institut f\"ur Kernphysik, P.O. Box 103980, D 69029
Heidelberg, Germany
\and
 Yerevan Physics Institute, 2 Alikhanian Brothers St., 375036 Yerevan,
Armenia
\and
Centre d'Etude Spatiale des Rayonnements, CNRS/UPS, 9 av. du Colonel Roche, BP
4346, F-31029 Toulouse Cedex 4, France
\and
Universit\"at Hamburg, Institut f\"ur Experimentalphysik, Luruper Chaussee
149, D 22761 Hamburg, Germany
\and
Institut f\"ur Physik, Humboldt-Universit\"at zu Berlin, Newtonstr. 15,
D 12489 Berlin, Germany
\and
LUTH, UMR 8102 du CNRS, Observatoire de Paris, Section de Meudon, F-92195 Meudon Cedex,
France
% ...
}

\date{Received month day, year; accepted month day, year}

\abstract
{
The blazar \object{1ES\,1101-232} was observed with the High Energy Stereoscopic System (H.E.S.S.) of Atmospheric Cherenkov Telescopes (ACT) 
in 2004 and 2005, for a live time of 43 hours.
VHE ($E$$>$$10^{11}$\,eV) $\gamma$-rays were detected for the first time from this object.
}
{
VHE
observations of blazars are used to investigate the inner parts of the blazar jets, 
and also to study the
extragalactic background light (EBL) in the near-infrared band.
}
{
Observations in 2005 were conducted in a multiwavelength campaign, together with the RXTE satellite and 
optical observations. In 2004, simultaneous observations with XMM-Newton were obtained.
}
{
1ES\,1101-232 was detected with H.E.S.S. with an excess of 
649
photons, at a significance of 
10\,$\sigma$.
The measured VHE $\gamma$-ray flux 
amounts to 
$\mathrm{d}N/\mathrm{d}E = (5.63\pm0.89)\times 10^{-13} (E/\mathrm{TeV})^{-(2.94\pm0.20)}
\mathrm{cm}^{-2}\mathrm{s}^{-1}\mathrm{TeV}^{-1}$, above a spectral energy threshold of 225\,GeV.
No significant variation of the VHE $\gamma$-ray flux on any time scale
was found.
1ES\,1101-232 exhibits a very hard spectrum, and
at a redshift of $z=0.186$, is the blazar with the highest confirmed redshift detected in VHE $\gamma$-rays so far.
}
{
The data allow the construction of truly simultaneous spectral energy distributions of the source, from the optical
to the VHE band.
Using an EBL model with $\nu F_{\nu}=14\,\mathrm{nWm^{-2}sr^{-1}}$ at $1.5\,\mathrm{\mu m}$ 
as presented in Aharonian et al. (2006a) suggests an intrinsic VHE power output peak 
of the source at above 3\,TeV.
}


\authorrunning{Aharonian et al.}
\titlerunning{VHE gamma-ray emission from 1ES\,1101-232 with H.E.S.S.}

\offprints{G.Puehlhofer@lsw.uni-heidelberg.de}

\maketitle

\section{Introduction}
\label{S:intro}

Blazars (BL Lacs and Flat Spectrum Radio Quasars) are thought to be active galactic nuclei (AGN) that have their jet axis oriented close to
the line of sight of the observer. 
The broadband spectral energy distribution (SED, $\nu F_\nu$ representation) of blazars is
characterized by two peaks, one at optical to X-ray energies, and another at $\gamma$-ray energies. 
The low-energy branch is commonly explained as electron synchrotron emission.
The high-energy branch 
can be explained in a variety of ways. In leptonic scenarios, it is assumed to result from 
Inverse Compton (IC) emission from the same electron population, up-scattering the self-generated synchrotron photons or external photons 
(synchrotron self-Compton, SSC, and external Compton, EC, 
e.g., \citealt{mushotzky1977sscinagn,madejski1983ssceinstein,ghisellinietal1985ssc,band1986ecfields,dermer1993ecblazarmodel,
sikora1994ecblazarmodel}).
In alternative hadronic scenarios, accelerated protons are the main source of high-energy radiation,
either directly or through the production of secondary particles
\citep[e.g.,][]{biermann1987protonjet,
aharonian2000protonjet,pohl2000protonjet,muecke2001protonjet}. 
Observationally, 
the SED and the variability in the different bands
carry 
the
information about the acceleration processes at work in the jet, and could ultimately
also shed light on the energy transfer mechanism of the central engine -- a supermassive black hole -- into the jet.

Over the past fourteen years, VHE $\gamma$-ray emission from
approximately a dozen
blazars has been detected 
\citep[see, e.g.,][for a recent review]{ong2005icrcreview}.
Both the detection of fast variability and the availability of broadband observations
-- especially including X-ray measurements -- have been
used to constrain individual source parameters \citep[e.g.,][]{krawczynski2002501model,aharonian20052155mwl}.
Increasing the number of known {\em VHE blazars}, especially at higher redshift,
is 
of importance for two reasons:

(1) Relatively little is still known about the average behaviour of VHE blazars. 
Most VHE blazars detected so far belong to the classes of X-ray selected BL Lacs (XBL) or high-frequency peaked BL Lacs (HBL),
but population studies are restricted by 
the low number of sources.
Previous detections of VHE blazars have also been biased towards high states of the sources,
because of the limited sensitivity of the available instruments. 
It was shown only recently that quiescent states can be detected 
now in short ($\sim$hours) observations \citep{aharonian20052155mwl}.
Little is 
known about average activity cycles and flare time scales,
except for a few sources:
\object{Mkn\,421} \citep[e.g.,][]{aharonian2002hegra421,blazejowski2005whipple421}, 
\object{Mkn\,501} \citep[e.g.,][]{aharonian1999hegra501,krawczynski2002501model},
and \object{1ES\,1959+650} \citep[e.g.,][]{krawczynski2004orphan1959,albert2006magic1959}.

(2) Source photons above $\sim$100\,GeV are attenuated by the EBL through $\gamma$-$\gamma$-interactions.
Therefore, an EBL density in the 
relevant waveband range
(typically $\sim$0.1 to $\sim$10$\,\mathrm{\mu m}$) 
has to be assumed to derive the intrinsic blazar spectrum. 
Conversely, if it is possible to determine or constrain the intrinsic blazar spectrum through models,
then constraints on the EBL density in the respective waveband can be obtained
\citep[e.g.][hereafter AHA06a]{coppi1999ebltev,aharonian2006ebl}.

To date, 1ES\,1101-232 is the most distant VHE
blazar known with confirmed redshift ($z=0.186$).  
It should be noted that the hard spectrum observed from 1ES\,1101-232 and its relatively large redshift allow for strong constraints on the EBL density.
This is described in detail in a separate paper (\citeauthor*{aharonian2006ebl}).

In this paper, the discovery of VHE $\gamma$-ray emission from the blazar 1ES\,1101-232 with H.E.S.S. is reported. 
The paper is organized as follows: In Section \ref{S:sourcedesc}, the source characteristics of 1ES\,1101-232 are 
described.
The results of the H.E.S.S. observations of 1ES\,1101-232 are presented in Section \ref{S:vheobs}.
In Section \ref{S:broadband}, 
multifrequency observations that were performed contemporaneously 
to the H.E.S.S. observations
are reported.
In Section \ref{S:discussion},
we concentrate
on the interpretation of the
spectral energy distribution 
derived from the source.

\section{The HBL object 1ES\,1101-232}
\label{S:sourcedesc}

Emission from 1ES\,1101-232 was first detected by the Ariel-5 X-ray satellite, the source \object{A\,1059-22} was however
misidentified with the \object{Abell 1146} cluster of galaxies at z=0.139 at that time 
\citep{mchardy1981arieldetection,maccagni1978misidentification}.
The HEAO-1 source \object{H\,1101-232} was later
correctly identified as a BL Lac type object, using the optical and radio counterparts 
\citep{buckley1985opticalidentification,remillard1989opticalidentification}.
The source has been detected by multiple X-ray observatories, and
for the purpose of this paper, the commonly used name 1ES\,1101-232 from the Einstein slew survey is adopted
\citep{elvis1992einsteinsurvey,perlman1996einsteinbllac}.

1ES\,1101-232 resides in an elliptical host galaxy
at a redshift of $z=0.186$ \citep{remillard1989opticalidentification,falomo1994opticalspectra}.
The host is presumably part of a galaxy 
cluster \citep{remillard1989opticalidentification,pesce1994nearbygalaxies}.
VLA maps of the BL Lac show a one-sided diffuse structure to the north of $\sim45''$ size,
but no well-collimated jet outside a few kpc distance from the core \citep{laurentmuehleisen1993vla}.
In the optical, the host galaxy is resolved \citep{remillard1989opticalidentification,abraham1991optical}.
\citet{falomo1993opticalir} deduced its brightness using a spectroscopic deconvolution of host and BL Lac.
The most recent estimate of $m_\mathrm{R}=16.41$ was derived from an angular profile fit \citep{falomo2000optical}.
The galaxy
is one of the brightest BL Lac host galaxies so far detected, and also 
the only one known with significantly boxy isophotes \citep{falomo2000optical},
indicating a merger process or extra dust components.
The BL Lac itself has a typical brightness of $m_V=16\textnormal{--}17$ \citep[e.g.,][]{remillard1989opticalidentification}.
The optical emission from 1ES\,1101-232 has typically varied on the timescale of months \citep[e.g.,][]{remillard1989opticalidentification}.
Optical flares on intraday timescales 
have
also been claimed in one observation
\citep{romero1999microvariability}.

The source has been classified earlier as an XBL \citep[e.g.,][]{scarpa1997opticalspectra},
and later on as an HBL \citep[e.g.,][]{donato2001blazarsequence}, because of the dominance of synchrotron emission
in the X-ray band.
Several authors have concluded from the broadband characteristics of 1ES\,1101-232 that this source is
expected to 
emit VHE $\gamma$-ray emission at flux levels detectable by instruments like H.E.S.S. \citep[e.g.,][]{wolter20001101,costamante2002tevcandidate}.
Previous VHE observations with the Durham Mark 6 telescope in 1998 have
only yielded flux limits \citep{chadwick1999limits}.
Also, in the GeV $\gamma$-ray domain, EGRET did not detect emission from 1ES\,1101-232 \citep{lin1996egret}.

In previously published SSC models \citep{wolter20001101,costamante2002tevcandidate}, the IC peak was 
generally expected to be around $100\,\mathrm{GeV}$,
but this seems not to be the case as 
shown in this paper.

\section{H.E.S.S. observations of 1ES\,1101-232}
\label{S:vheobs}

Observations were made with the
H.E.S.S. Cherenkov telescopes in April and June 2004, and in March 2005. 
On June 8$^{\mathrm{th}}$ 2004, 
also
XMM-Newton X-ray observations
were obtained,
scheduled such that simultaneous H.E.S.S. observations could be conducted.
Following the detection of a weak signal in the 2004 H.E.S.S. observations,
an extended multifrequency
campaign was organized
for 11 nights in March 2005, to study the broadband emission from 
1ES\,1101-232 and to search for 
(possibly correlated) variability in the different wavebands. 
Simultaneous observations were carried out with H.E.S.S.,
X-ray measurements with RXTE, and optical measurements with the ROTSE 3c robotic telescope.

\subsection{The H.E.S.S. experiment}

The H.E.S.S.\ experiment \citep{hinton2004hess} consists in phase I of four
ACTs operating stereoscopically.
Each telescope consists of a tesselated 13\,m diameter ($107$\,m$^2$
surface area) mirror, which focuses the Cherenkov light from
the showers of secondary particles created by the interaction of
$\gamma$-rays in the atmosphere onto a camera in the focal plane.
Each camera consists of 960 photomultipliers with a pixel size of
$0.16^\circ$, providing a field of view (FoV) of $5^\circ$. 
The array is located in the Khomas highlands in Namibia 
($-23^\circ16'$, $16^\circ30'$, 1835\,m a.s.l.).

The angular resolution of the stereo system is better than 0.1$^\circ$ per event.
The energy threshold of H.E.S.S. is about 100\,GeV (at zenith), with spectral
measurements possible above $\sim$150\,GeV with an energy resolution of
15\%. The energy threshold increases with zenith angle. For the data set of 1ES\,1101-232 
discussed in this paper, the time-averaged spectrum presented in this paper has 
an energy threshold of 225\,GeV.

The minimum detectable point source flux 
above 100\,GeV with H.E.S.S. is $\sim 4 \times 10^{-12}\mathrm{erg\,cm^{-2}s^{-1}}$ 
for a $5\,\sigma$ detection in 25 hours, corresponding to $\sim$10\,mCrab \citep{aharonian2006crab}.
The sensitivity enabled a $\ge 3\,\sigma$ detection per night ($\sim 5$ hours) 
in the case of the 1ES\,1101-232 observations.

\subsection{H.E.S.S. observations of 1ES\,1101-232}

First observations of 1ES\,1101-232 with H.E.S.S. were performed during four nights in
April 2004, for a total live time of 2.7 hours after quality selection, and for six nights in June 2004, 
for a total of 8.4 hours after quality selection. The total data set in March 2005 after quality selection amounts to 31.6 hours live time. 
The observation log is shown in Tab.\,\ref{T:hessobslog}.

H.E.S.S. observations were taken in runs of typically 28\,min each. Runs were taken in {\it wobble mode}\footnote{
In wobble mode,
the source is
displaced with respect to the center of the FoV, 
the sign of the offset alternating between consecutive runs.
This 
permitted 
continuous monitoring of 
1ES\,1101-232.
}, with a declination or right ascension
offset of
$\pm0.5\degr$ in 2004 and $\pm0.7\degr$ in 2005.
Since the $\gamma$-ray acceptance of the H.E.S.S. instrument is nearly constant within $1\degr$ radius from the FoV center,
this change of observation mode 
results in a slight sensitivity increase,
as a larger background control area with equal acceptance in the FoV can be used.

The data cleaning to derive a set of good
quality
runs, which are used in the data analysis, consists
of two filtering processes. 
First, from the individual shower images as recorded from triggered telescopes, all pixels that 
have not worked properly are removed; occasionally, entire cameras have to be excluded from individual runs. 
Then runs are discarded that show a too low or fluctuating event trigger rate,
caused by bad atmospheric conditions or clouds.

\begin{table} 
\begin{center}
  \begin{tabular}{l|l|l|l|l|l} 
\hline
 Period & $^1$MJD & $^2T$ (runs)& $^3F$                 & $^4S$ & X-ray \\ 
\hline
 Apr        & 53111 & 0.91 ( 3) & $ 1.1^{+ 2.6}_{ 1.1}$ & 1.63  &     \\ 
 2004       & 53113 & 0.71 ( 2) & $ 6.5^{+ 3.7}_{ 3.6}$ & 0.90  &     \\ 
            & 53115 & 0.83 ( 2) & $ 3.0^{+ 2.7}_{ 2.2}$ & 0.33  &     \\ 
            & 53117 & 0.21 ( 1) & $ 5.7^{+ 6.4}_{ 4.5}$ & 1.82  &     \\ 
\hline
 Jun        & 53162 & 0.85 ( 2) & $ ^5 <4.6           $ & 0.41  &     \\ 
 2004       & 53163 & -         &                       &       &     \\ 
            & 53164 & -         &                       &       &     \\ 
            & 53165 & 3.18 ( 7) & $ 5.5^{+ 2.6}_{ 2.6}$ & 3.20  & XMM \\ 
            & 53166 & 2.72 ( 6) & $ 8.2^{+ 2.8}_{ 2.8}$ & 3.36  &     \\ 
            & 53167 & 1.72 ( 5) & $ 4.6^{+ 2.4}_{ 2.4}$ & 1.81  &     \\ 
\hline
 Mar        & 53435 & -         &                       &       & XTE \\
 2005       & 53436 & 5.15 (11) & $ 7.7^{+ 3.6}_{ 3.6}$ & 5.63  & XTE \\ 
            & 53437 & 5.29 (12) & $ 2.1^{+ 3.1}_{ 1.8}$ & 2.87  & XTE \\ 
            & 53438 & 5.12 (10) & $ 5.4^{+ 1.7}_{ 1.7}$ & 5.00  & XTE \\ 
            & 53439 & 5.01 (10) & $ 4.6^{+ 1.6}_{ 1.6}$ & 3.39  & XTE \\ 
            & 53440 & 3.25 ( 7) & $ 4.2^{+ 2.1}_{ 2.1}$ & 3.10  & XTE \\ 
            & 53441 & 1.65 ( 3) & $ 1.6^{+ 2.7}_{-1.6}$ & 2.16  & XTE \\ 
            & 53442 & -         &                       &       & XTE \\ 
            & 53443 & 2.42 ( 5) & $ 5.0^{+ 2.0}_{ 2.0}$ & 2.47  & XTE \\ 
            & 53444 & 1.80 ( 4) & $ 8.0^{+ 2.6}_{ 2.6}$ & 3.59  & XTE \\ 
            & 53445 & 1.92 ( 4) & $ 2.4^{+ 1.9}_{ 1.6}$ & 1.90  & XTE \\ 
\hline
  \end{tabular}
\end{center}
\caption{Log of the H.E.S.S. observations on 1ES\,1101-232 in 2004 and 2005. Numbers reflect the observations
after data cleaning and good run selection. Nights with observations performed on 1ES\,1101-232, where however 
all H.E.S.S. observations needed to be discarded because of weather selections, are marked with a `-'.
$^1$Modified Julian date.
$^2$Live time $T [\mathrm{hours}]$. 
$^3$Flux $F(E$$>$$200\,\mathrm{GeV}) [10^{-12}\mathrm{erg}\,\mathrm{cm}^{-2}\mathrm{s}^{-1}]$. 
$^4$Detection significance $S$ in units of standard deviations.
$^5$Upper limit at 99\% confidence level.
}
  \label{T:hessobslog}
\end{table}

\begin{figure}
\centering
  \includegraphics[width=8.7cm]{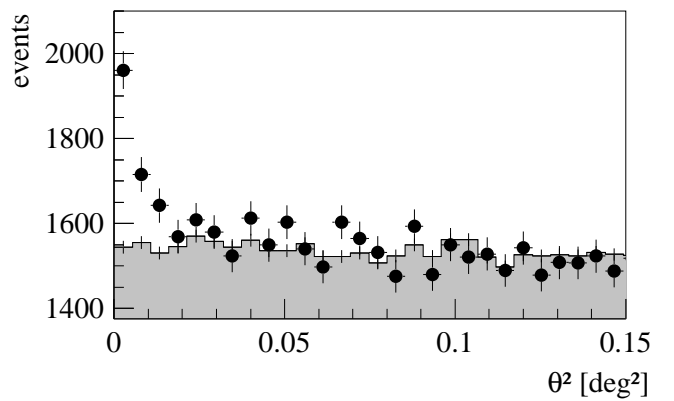}
\caption{Angular event distribution. 
Events are from the entire H.E.S.S. data set on 1ES\,1101-232, after shape cuts to suppress
a large fraction of the background. 
The filled circles denote the 
event distribution in squared distance with respect to
the direction to 1ES\,1101-232.
The filled histogram represents the expected 
background.
For this histogram, the distributions obtained from seven control regions were summed and renormalized.
 }
  \label{F:1101thetasquared}
\end{figure}

\begin{table}[h]
\begin{center}
  \begin{tabular}{lll} 
\hline
\multicolumn{3}{l}{Cuts} \\
&image distance to FoV center  & $<2\,\mathrm{deg}$    \\
&image amplitude               & $>80\,\mathrm{ph.e.}$ \\
&number of telescopes          & $\ge2$                \\
&cut on shower width           & $-2.0<mrsw<0.9$\\
&cut on shower length          & $-2.0<mrsl<2.0$\\
&distance to source $\theta^2$ & $<0.0125\,\mathrm{deg}^2$\\ 
\hline
\multicolumn{3}{l}{Total Data Set (April 2004, June 2004, March 2005)} \\ 
&$N_{\mathrm{on}}$ (events)    & 4276\\
&$N_{\mathrm{off}}$ (events)   & 54345 \\
&normalisation $\alpha=A_{\mathrm{on}}/A_{\mathrm{off}}$        & 0.0667 \\
&excess (events)               & 649.0\\
&significance                  & 10.1 $\sigma$\\
\hline
\multicolumn{3}{l}{March 2005 Data Set} \\
&$N_{\mathrm{on}}$ (events)    & 3028\\
&$N_{\mathrm{off}}$ (events)   & 42427\\
&normalisation $\alpha=A_{\mathrm{on}}/A_{\mathrm{off}}$        & 0.0597 \\
&excess (events)               & 495.8\\
&significance                  & 9.3 $\sigma$\\
\hline
\multicolumn{3}{l}{June 2004 Data Set} \\
&$N_{\mathrm{on}}$ (events)    & 902\\
&$N_{\mathrm{off}}$ (events)   & 8531\\
&normalisation $\alpha=A_{\mathrm{on}}/A_{\mathrm{off}}$        & 0.0926 \\
&excess (events)               & 112.4\\
&significance                  & 3.7 $\sigma$\\
\hline
\multicolumn{3}{l}{April 2004 Data Set} \\
&$N_{\mathrm{on}}$ (events)    & 346\\
&$N_{\mathrm{off}}$ (events)   & 3405\\
&normalisation $\alpha=A_{\mathrm{on}}/A_{\mathrm{off}}$        & 0.09091 \\
&excess (events)               & 36.5\\
&significance                  & 1.9 $\sigma$\\
\hline
  \end{tabular}
\end{center}
\caption{H.E.S.S. analysis parameters and event numbers. Values are given for the total H.E.S.S. data set as well as for the 
three data subsets considered.
}
  \label{T:hessanalysisnumbers}
\end{table}

\subsection{Analysis of the H.E.S.S. data}
\label{SS:HESSanalysis}

The H.E.S.S. data have been processed and analysed according to the standard analysis chain as detailed for instance in
\citet{aharonian2005standardanalysis} and \citet{benbow2005standardanalysis}. Shower images as recorded by individual telescopes are parametrized using the Hillas
parametrisation. The shower direction (i.e., the origin of the incoming $\gamma$-ray) and the projected 
shower impact point on the ground are
determined from a stereoscopic reconstruction of the air shower. 
A large fraction of the background is suppressed using cuts on the shower shape parameters
mean reduced scaled width $mrsw$ and length $mrsl$.
Photon directions 
are used for
a one-dimensional projection including background estimate ($\theta^{2}$-plot, Fig.\,\ref{F:1101thetasquared})
and to reconstruct the source location (Fig.\,\ref{F:1101dss}).
The 
background after shape cuts is estimated from different control regions in the FoV, having the same distance to the
center of the FoV in order to be independent of the radial acceptance change \citep{hinton2005background}. 
For this data set, a background region with an area
$A_{\mathrm{off}}$ 11 (in 2004) or 15 (in 2005) times larger than the on-source area $A_{\mathrm{on}}$ was used.
The applied standard cuts (see Tab.\,\ref{T:hessanalysisnumbers}) were optimized on a simulated source with 10\% of the flux from the Crab Nebula and a
Crab-like power-law spectrum $\propto E^{-\Gamma}$ with $\Gamma$=$2.6$.

Energy estimates for individual photons are based on the comparison of the measured image amplitudes to data from simulated events, using the
measured shower core position and zenith angle as parameters. The resolution per event is $\sim$15\% above the spectral
threshold, which is defined as the energy threshold 
above which the energy reconstruction bias is less than 10\%. Only events above this
safe threshold are used to compute energy spectra. The effective area is based on Monte Carlo simulations and is depending on zenith
angle and system configuration (i.e., which telescopes are included in a particular run).

Compared to the analysis results used in \citeauthor*{aharonian2006ebl}, an improved energy calibration of the telescope system
was applied to the data, better taking into account the long-term optical sensitivity changes of the instrument \citep{aharonian2006crab}.
For the given total data sample, 
this
energy scale recalibration yields a safe energy threshold of 225\,GeV 
(compared to 165\,GeV used in \citeauthor*{aharonian2006ebl})
and a flux normalisation increase of 27\% at 1\,TeV. 
After this correction, the systematic flux uncertainty is now estimated as 
20\% \citep{aharonian2006crab}. Reconstructed spectral indices were not affected significantly by these calibration updates,
the systematic error estimate for reconstructed photon indices is $\Delta \Gamma_{\mathrm{sys}} \sim 0.1$
(\citeauthor*{aharonian2006ebl}; \cite{aharonian2006crab}).
The recalibration slightly increased the background noise in this data set which therefore has an excess significance
of $10.1\,\sigma$, slightly smaller than the detection significance of $11.7\,\sigma$ that was derived from the original data set used in
\citeauthor*{aharonian2006ebl}.

The results derived have been verified using a completely independent calibration and analysis chain, 
which is described for instance in \citet{aharonian20052356} and \citet{lemoine20063dmodel}.

\subsection{Skymap and identification of the VHE $\gamma$-ray source}

\begin{figure}
  \includegraphics[width=8.7cm]{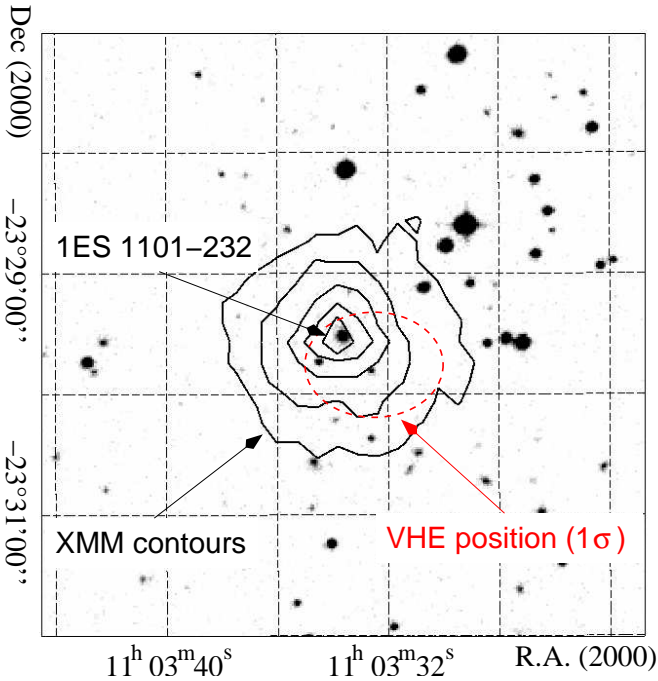}
\caption{Sky map of the region surrounding 1ES\,1101-232.
An R-band image 
made by the Anglo-Australian observatory with the UK Schmidt telescope
is shown in grey-scale.
The host galaxy (labeled 1ES\,1101-232 in the image) of the BL Lac is resolved as an elliptical galaxy, with boxy isophotes at larger radii \citep{falomo2000optical}.
The dashed ellipse denotes the $1\,\sigma$ error
of the reconstructed VHE $\gamma$-ray source position. 
The solid contours are
from the XMM-Newton X-ray measurements with the MOS\,2 camera, as discussed in Section \ref{SS:xmmobs}.
The contour spacing is logarithmic (0.1, 0.3, 1.4, 5.8 and 24\% of the peak intensity), the outermost contour
also denotes the edge of the effective MOS\,2 FoV for this small window mode observation.
}
  \label{F:1101dss}
\end{figure}

Source confusion is generally a minor issue in VHE data, especially from extragalactic sources. 
No other source near 1ES\,1101-232 ($11^{\mathrm{h}}03^{\mathrm{m}}37.57^{\mathrm{s}},-23^{\circ}29'30.2''$, J2000, \citealt{laurentmuehleisen1993vla}) 
is a good candidate for VHE $\gamma$-ray emission.
The VHE $\gamma$-ray source location was derived from the skymap of VHE photons and is
$11^{\mathrm{h}}03^{\mathrm{m}}36.5^{\mathrm{s}} \pm 2.5^{\mathrm{s}}_{\mathrm{stat}}, -23^{\circ}29'45'' \pm 26''_{\mathrm{stat}}$ (J2000), 
which is consistent with the above given radio position, and also with the optical and X-ray positions as shown in Fig.\,\ref{F:1101dss}.
As the present VHE data do not exhibit variability, 
the identification of the VHE $\gamma$-ray source is 
based on its location 
and the interpretation of the SED.

Figure \ref{F:1101dss} also shows that the X-ray imaging data,
which were 
obtained with XMM-Newton, are compatible with the XMM-Newton point
spread function. 
This excludes the influence of possible nearby X-ray sources in the FoV of non-imaging X-ray measurements such as RXTE.

\subsection{VHE $\gamma$-ray light curve}

\begin{figure}
  \includegraphics[width=8.7cm]{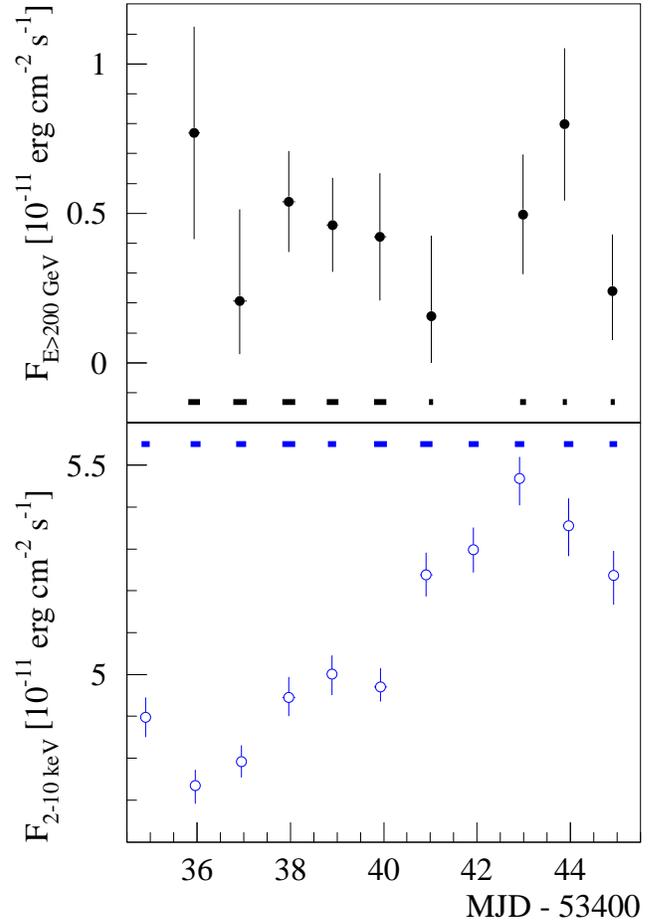}
\caption{March 2005 light curve of 1ES\,1101-232. MJD=53435 corresponds to the night Mar.\,5.-6.
{\bf Upper panel:} VHE $\gamma$-ray flux as measured with H.E.S.S., in nightly averaged bins. Within errors, the flux was constant.
{\bf Lower panel:} X-ray flux ($2\textnormal{--}10\,\mathrm{keV}$), derived from the PCU\,2 detector onboard RXTE.
Note that the flux scale is truncated,
the X-ray flux difference from minimum to maximum is $\sim$15\%.
The thick horizontal bars in both panels denote the times (first to last measurement of an observing night) 
when the VHE and X-ray data were taken, indicating the high degree of simultaneity of the two data sets.
}
  \label{F:1101hessxtelcmar2005}
\end{figure}

The fluxes derived from the three H.E.S.S. data sets (April 2004, June 2004, March 2005) 
are statistically compatible with each other, indicating that the VHE $\gamma$-ray flux
has remained constant throughout these observing periods. However, as the detections from the
April ($1.9\,\sigma$) and June 2004 ($3.7\,\sigma$) data sets alone are not very significant,
only flux variations of a factor of approximately two or larger can be excluded. 
Night-by-night variations were also searched for, but the 
corresponding light curve is compatible with a constant flux
(probability for a constant flux of the total data set 85\%, of the March 2005 data set 64\%). 

In Fig.\,\ref{F:1101hessxtelcmar2005}, the nightly averaged light curve of H.E.S.S. is shown for the March 2005 period,
together with simultaneous X-ray measurements from the RXTE satellite.
The VHE $\gamma$-ray flux is given in integral flux units above 200\,GeV and was computed under the assumption of the
time-averaged spectral index of $\Gamma=2.94$. The X-ray fluxes were similarly derived, details are given in Section \ref{SSS:xtexray}.
The RXTE data indicate an X-ray flux variation of $\sim$15\% (min-max), whereas the 
simultaneously taken 
H.E.S.S. data are not sensitive enough to detect possible correlations with similar amplitudes
in the X-ray and VHE bands.

\subsection{VHE $\gamma$-ray spectrum}
\label{SS:vhespectrum}

The differential energy spectra derived for the entire H.E.S.S. data set, as well as for the June 2004 and the March 2005 data subsets,
are shown in the left panels of Fig.\,\ref{F:1101hessspec}. 
The low statistics of the April 2004 data set prevented us from performing a spectral analysis on that subset.
The measured spectra are compatible with power laws,
Tab.\,\ref{T:hessspectranumbers} lists the corresponding photon indices $\Gamma_{\mathrm{abs}}$ derived from fits between 0.2 and 4\,TeV. 
The fit for the total
spectrum gives
$\mathrm{d}N/\mathrm{d}E = (5.63\pm0.89)\cdot 10^{-13} (E/\mathrm{TeV})^{-(2.94\pm0.20)}
\mathrm{cm}^{-2}\mathrm{s}^{-1}\mathrm{TeV}^{-1}$ above a spectral energy threshold of 225\,GeV.
The integral flux is $F(E>200\,\mathrm{GeV}) = (4.5 \pm 1.2) \cdot 10^{-12}\mathrm{erg\,cm}^{-2}\mathrm{s}^{-1}$.

Spectral bins above 4\,TeV do not contain a significant flux detection. Therefore, from these bins 99\% upper limits were derived
and shown in Fig.\,\ref{F:1101hessspec}.
The photon indices $\Gamma_{\mathrm{abs}}$ were derived excluding these bins.
Table \ref{T:hessspectranumbers} additionally lists photon indices $\Gamma^{*}_{\mathrm{abs}}$ from single power law fits,
for which these flux estimates above 4\,TeV were included in $\chi^{2}$-fits.
This is a viable procedure since the error bars are dominated by background estimates and are therefore mostly Gaussian.
However, the results depend 
on the assumption that the power laws would extend to these high energies.

\subsection{VHE spectrum corrected for EBL absorption}
\label{SS:vhespectrumdeabsorbed}

In the following, the data sets were corrected for EBL absorption, to investigate whether the intrinsic spectra show
evidence for a cut-off towards high energies. 
Spectral changes between periods were also searched for.
The deabsorbed spectra are used in the SED, as discussed in Section \ref{S:discussion}.

The intrinsic
spectrum of 1ES\,1101-232 depends on the assumed EBL spectrum. For a detailed discussion about the EBL absorption
of 
VHE $\gamma$-ray spectra, we refer the reader to, e.g., \citet{aharonian2001proceedings}.
Corrections with a range of plausible EBL spectra result in deabsorbed spectra 
of 1ES\,1101-232 that can be described over the entire detected energy range,
i.e. between 0.2 and 4\,TeV,
by a single power law of photon index $\Gamma_{\mathrm{deabs}}$,
i.e., $\Gamma_{\mathrm{deabs}} = \Gamma_{\mathrm{abs}} - \Delta\Gamma$, 
see \citeauthor*{aharonian2006ebl}; \citet{stecker2006deltagamma}.

\subsubsection{{\em Maximum} EBL}
\label{SSS:maximumebl}

\begin{figure*} 
\centering
  \includegraphics[width=11cm]{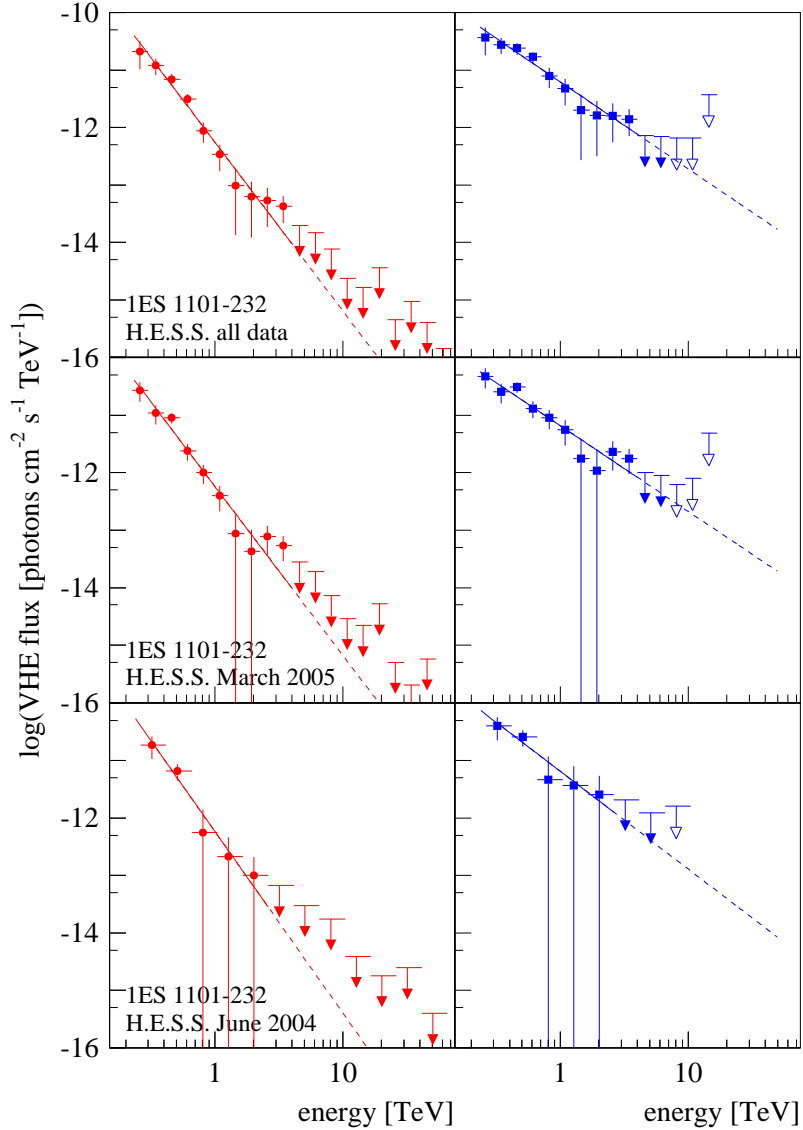}
\caption{
VHE $\gamma$-ray spectra from 1ES\,1101-232. 
{\bf Left panels:} Reconstructed photon flux, as measured with H.E.S.S.
{\bf Right panels:} Photon spectra after correction for maximum EBL absorption, 
using an EBL model with $14\,\mathrm{nWm^{-2}sr^{-1}}$ at $1.5\,\mathrm{\mu m}$ as described in Section \ref{SSS:maximumebl}. 
Upper limits in these deabsorbed spectra at energies above 7\,TeV are shown as open symbols only, 
because of strong EBL uncertainties at these high energies.
In all panels, solid lines denote 
power-law fits between 0.2 and 4\,TeV.
Extrapolations of these power laws to higher energies are shown as dashed lines.
}
  \label{F:1101hessspec}
\end{figure*}

\begin{table*} 
  \begin{center}
  \begin{tabular}{l||c|c||c|c} 
    \hline
                 & $\Gamma_{\mathrm{abs}}$  & $\Gamma^{*}_{\mathrm{abs}}$ & $\Gamma_{\mathrm{deabs}}$ & $\Gamma^{*}_{\mathrm{deabs}}$ \\ 
                 & \small{$0.23\textnormal{--}4.0\,\mathrm{TeV}$}               & \small{$0.23\textnormal{--}16.9\,\mathrm{TeV}$}                         & \small{$0.23\textnormal{--}4.0\,\mathrm{TeV}$}                & \small{$0.23\textnormal{--}7.1\,\mathrm{TeV}$}  \\ \hline 
      All Data   & $ 2.94^{+0.20}_{-0.21}$  & $ 3.10^{+0.17}_{-0.19}$             & $ 1.51^{+0.17}_{-0.19}$   & $ 1.74^{+0.15}_{-0.17}$\\
      March 2005 & $ 2.94^{+0.21}_{-0.23}$  & $ 3.08^{+0.18}_{-0.21}$             & $ 1.49^{+0.19}_{-0.20}$   & $ 1.68^{+0.16}_{-0.18}$\\
      June 2004  & $ 3.16^{+0.48}_{-0.61}$  & $ 3.45^{+0.41}_{-0.59}$             & $ 1.70^{+0.47}_{-0.61}$   & $ 2.19^{+0.40}_{-0.60}$\\
    \hline
  \end{tabular}
  \end{center}
\caption{Photon indices from power-law fits to the VHE spectra of 1ES\,1101-232.
$\Gamma_{\mathrm{abs}}$ and $\Gamma^{*}_{\mathrm{abs}}$ are from fits to the measured spectra, $\Gamma_{\mathrm{deabs}}$
and $\Gamma^{*}_{\mathrm{deabs}}$ from fits to the deabsorbed spectra as described in Section \ref{SSS:maximumebl}. 
$\Gamma_{\mathrm{abs}}$ and $\Gamma_{\mathrm{deabs}}$ correspond to the fits shown as solid lines in Fig.\,\ref{F:1101hessspec}.
Photon indices $\Gamma^{*}_{\mathrm{abs}}$ and $\Gamma^{*}_{\mathrm{deabs}}$ are from fits including spectral bins above 4\,TeV, which are 
compatible with zero flux, under the assumption that the power laws extend to these high energies.
}
  \label{T:hessspectranumbers}
\end{table*}

To represent what 
\citeauthor*{aharonian2006ebl} considered as the highest possible EBL level,
the EBL shape {\em P0.45} 
from
\citeauthor*{aharonian2006ebl} was used 
(cf. also Section \ref{SS:vheebllimit}). 
This shape reflects an EBL level of $14\,\mathrm{nWm^{-2}sr^{-1}}$ at $1.5\,\mathrm{\mu m}$ after scaling down by 15\%
to take galaxy evolution effects into account. 
The such deabsorbed spectra of 1ES\,1101-232 are shown in the right panels of Fig.\,\ref{F:1101hessspec}.
The 
fit to the
deabsorbed spectrum of the total data set 
in the energy range 0.2 to 4\,TeV
yields 
$\Gamma_{\mathrm{deabs}}=1.51\pm0.17$.
The spectra of the two subsets considered ($\Gamma_{\mathrm{Jun\,2004}}=1.70\pm0.47$ and 
$\Gamma_{\mathrm{Mar\,2005}}=1.49\pm0.19$)
are statistically compatible with each other.

Similar results are obtained when using the EBL ``baseline'' model described in 
\citet{stecker2006ebl}, as shown by \citet{stecker2006deltagamma}. Also the EBL model by \citet{primack2001ebl} 
yields similar numbers, after a slight reduction of 15\% to match the above quoted level of $14\,\mathrm{nWm^{-2}sr^{-1}}$.

It is of interest to test whether the
upper limits above 4\,TeV are compatible with a power-law extrapolation
of the lower energy spectra, or are indicative of a steepening of the spectrum. For the intrinsic spectra, this would imply an 
observational hint for a peak in the SED at that energy range. 
For the deabsorbed spectra, only 
those bins above 4\,TeV
could be included 
where the
chosen
EBL parametrisation ({\em P0.45})
can still be considered valid,
which restricts the range to $\leq 7\,\mathrm{TeV}$. 
As justified in Section \ref{SS:vhespectrum}, $\chi^{2}$-tests were applied.
For example, for the March 2005 spectrum, the probability of the spectrum being compatible with $\Gamma=1.49$
changes from 65\% to 49\%. 
Hence, there is no observational hint for a steepening
of the spectrum above $\sim 4\,\mathrm{TeV}$, but the data do not exclude a spectral change above that energy either.
Photon indices $\Gamma^{*}_{\mathrm{deabs}}$ from power-law
fits in the energy range 0.2 to 7\,TeV are listed in Tab.\,\ref{T:hessspectranumbers}.

\subsubsection{Further EBL levels}

Lowering the EBL level used for the deabsorbtion leads to a softening of the spectra.
Lower limits in the relevant EBL waveband range come from galaxy counts \citep{madau2000galaxycounts}
and are of the order of $10\,\mathrm{nWm^{-2}sr^{-1}}$ at $1.5\,\mathrm{\mu m}$. 
The resolved galaxy counts may however represent 
a too low EBL level because of possibly missed light \citep{totani2001ebl}.
Nevertheless, to represent the such constrained minimum EBL, 
the representation {\em P0.40} \citep{aharonian2006ebl} was chosen 
and scaled 
down by 15\% to take galaxy evolution effects into account.
Using this minimum EBL to deabsorb the spectra
result in $\Gamma_{\mathrm{deabs}}=1.85\pm0.18$, $\Gamma_{\mathrm{Jun\,2004}}=2.05\pm0.56$ and 
$\Gamma_{\mathrm{Mar\,2005}}=1.84\pm0.20$.

EBL models higher than the {\em maximum} EBL level were described, e.g., by \citet[][``fast evolution'' case]{stecker2006ebl} and by 
\citet[][``best fit" model]{kneiske2004ebl}, with an EBL density of about
$\nu F_{\nu}(1.5\,\mathrm{\mu m}) \simeq 20\,\mathrm{nWm^{-2}sr^{-1}}$.
As shown in \citet{stecker2006deltagamma}, the ``fast evolution'' EBL would lead to an intrinsic
spectrum with $\Gamma_{\mathrm{deabs}}\simeq1.0$. This result would be in conflict with the assumption 
of a limit on the intrinsic hardness of VHE blazar spectra, see next Section and \ref{SS:ebl}.

\subsection{EBL limit}
\label{SS:vheebllimit}

Following the discussion in \citeauthor*{aharonian2006ebl}, 
we assume
that the intrinsic blazar spectrum 
did not have
a photon index
significantly harder than 1.5 (i.e., $\Gamma_{\mathrm{deabs}}<1.5$), 
taking the present observational and theoretical knowledge of VHE blazar spectra into account. 
Applied to the data from 1ES\,1101-232, this limit results in an upper limit of the EBL
density in the waveband range of $0.75\textnormal{--}3.5\,\mathrm{\mu m}$,
with a peak density of $(14\pm4)\,\mathrm{nWm^{-2}sr^{-1}}$ at $1.5\,\mathrm{\mu m}$ for typical EBL shapes
as reported in \citeauthor*{aharonian2006ebl}.

Given that the updated calibration of the data set (Section \ref{SS:HESSanalysis})  
involves a change of the energy calibration which is slightly larger than the systematic error of $\pm 15$\%
quoted in \citeauthor*{aharonian2006ebl}, 
the procedure described in that paper to derive the EBL upper limit was repeated.
The updated analysis constrains the total spectrum to a power law with $\Gamma=2.94$ between 0.23 and 4.0\,TeV,
whereas in \citeauthor*{aharonian2006ebl} $\Gamma=2.88$ between 0.16 and 3.3\,TeV was used.
The EBL limit derived with these updated numbers differs from the value in \citeauthor*{aharonian2006ebl}
by less than 1\%, well below any statistical uncertainties involved.
The upper limit for the EBL spectrum 
as given above is therefore confirmed.

The error quoted for the peak density mostly comes from the statistical error of the photon index derived from the
1ES\,1101-232 data set. This uncertainty is therefore well represented by the statistical errors of the deabsorbed spectra 
of 1ES\,1101-232 discussed in the previous Section.
It is therefore inappropriate to translate the value
of $(14\pm4)\,\mathrm{nWm^{-2}sr^{-1}}$ into 
an
68\% or 95\% upper limit. We remark that -- because of the procedure described above --
the choice of $14\,\mathrm{nWm^{-2}sr^{-1}}$ yields by construction an intrinsic photon index of the total 1ES\,1101-232 data set of 1.5.

\section{Broadband observations of 1ES\,1101-232}
\label{S:broadband}

\subsection{The observation campaign in March 2005}

\subsubsection{X-ray data}
\label{SSS:xtexray}

110.2\,ksec RXTE observation time for monitoring observations
have been obtained, 
which were scheduled such that simultaneous H.E.S.S. observations
were possible for 11 consecutive nights in March 2005 (see Tab.\,\ref{T:hessobslog}). 
Because of Earth occultation and downtime in the southern Atlantic anomaly (SAA),
the average on-time of RXTE was 56\% during the ``nightly" satellite observation windows. 

RXTE data were analyzed using standard reduction routines. 
During most observations, two PCA detectors (PCU\,0 and PCU\,2) were active.
For the analysis presented here, 
only
results from PCU\,2 were used.
PCU\,0 has lost its front veto layer, and is therefore susceptible to unmodeled and unremoved background 
events
\citep{xue2005xte}.
Only the signal from the top layer (X1L, X1R) was used for optimum signal to noise ratio.
The \verb+STANDARD2+ data were extracted using \verb+XDF+ 
and processed using 
\verb+ftools+ \citep{blackburn1995ftools} from \verb+HEASOFT 6.0.4+.
The data were filtered using standard criteria recommended by the RXTE guest observer facility 
(avoidance of Earth's limb, SAA, and tracking offsets).
Electron contamination was suppressed by limiting the corresponding \verb+ELECTRON2+ parameter to below 0.1.
The effective exposure after all screening 
was 89.6\,ksec.
Background data were parametrized using \verb+pcabackest v3.0+, using the faint background model.
Response matrices were created with \verb+pcarsp v10.1+, and nightly and total spectra were extracted with \verb+saextrct v4.2d+. 
Spectral fitting was performed with \verb+XSPEC v11.3.1+, using PCA channels 5-32 (corresponding approximately to $3\textnormal{--}15\,\mathrm{keV}$). 
To account for Galactic absorption, a column density of $N_{\mathrm{H}}=5.76\times10^{20}\mathrm{cm^{-2}}$ was 
used in the spectral fitting 
(cf., e.g., \cite{wolter1998bepposax}, and also the \verb+PIMMS+ nH program). The influence of $N_{\mathrm{H}}$ is however marginal at 
this energy band.

No flux variability was found within any of the 11 observing nights. 
Between nightly averages, moderate changes were observed (the probability for a constant flux is $10^{-5}$, using the
nightly averaged count rates). 
No hint for spectral variability was found. 
The light curve shown in Fig.\,\ref{F:1101hessxtelcmar2005} was derived by fixing the spectral model to the broken power law derived for the
total spectrum (see next paragraph), while leaving the normalisation as the only free parameter.
Within statistical errors, the nightly fluxes derived are compatible with fluxes 
derived from single power-law fits with two free parameters (slope and normalisation).
We conclude that 1ES\,1101-232 was probably in a 
non-flaring
state during this
observation campaign. 
The simultaneous H.E.S.S. measurements are not sensitive to VHE $\gamma$-ray flux changes of 
similar variability amplitudes.

\begin{figure} 
  \includegraphics[width=8.7cm]{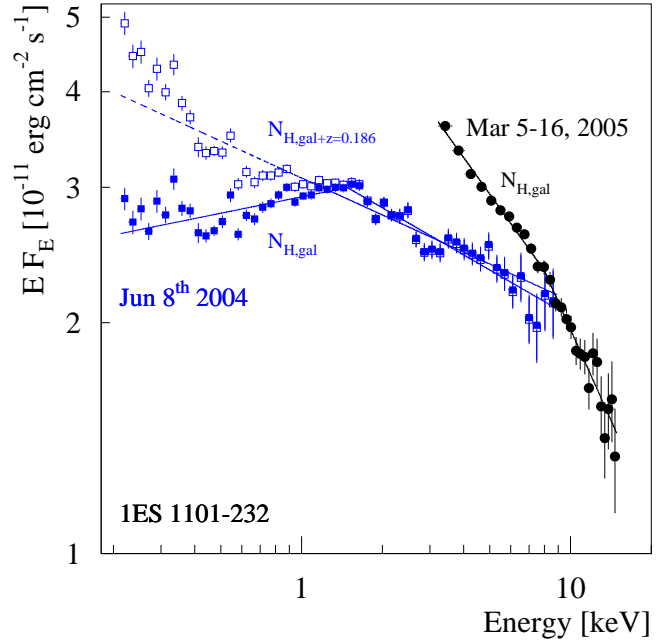}
\caption{Unfolded X-ray spectra from 1ES\,1101-232 in $E F(E)$-representation. 
Points between $3\textnormal{--}15\,\mathrm{keV}$ labeled 
Mar 5-16, 2005 are from RXTE, the line shows a broken power-law fit to the data. 
Points 
labeled Jun 8$^{\mathrm{th}}$, 2004 between $0.2\textnormal{--}10\,\mathrm{keV}$ are
from XMM-Newton. 
The 
filled 
squares
represent the reconstructed spectrum under the assumption of pure Galactic hydrogen absorption, and can be
fit by a broken power law.
For the spectrum shown with 
open 
squares,
in addition to Galactic also absorption in the source was allowed,
under the assumption of a pure power-law emission spectrum.}
  \label{F:1101xtespectra}
\end{figure}

As no spectral variability and only moderate flux changes were observed, 
a single averaged energy spectrum for the entire data set was derived.
The spectrum between 3 and 15\,keV is 
incompatible with a pure power law 
($\chi^{2}_{\mathrm{red}}= 2.08$ for 26 d.o.f., 
null-hypothesis probability\footnote{i.e., the probability that the assumed function fits the data} $10^{-3}$), 
but a broken power law yields an acceptable fit with 
$\chi^{2}_{\mathrm{red}}= 1.18$ (24 d.o.f., null-hypothesis probability 25\%).
The total unabsorbed flux is 
$F_{2\textnormal{--}10\,\mathrm{keV}} = (5.07\pm0.02_{\mathrm{stat}})\times10^{-11}\mathrm{erg}\,\mathrm{cm}^{-2}\mathrm{s}^{-1}$.
In Fig.\,\ref{F:1101xtespectra}, the unfolded spectrum is shown.
The spectrum is quite soft, with a photon index of $\Gamma=2.49\pm0.02_{\mathrm{stat}}$
below the break energy $E_{\mathrm{break}}=(7.9\pm1.0)\,\mathrm{keV}$,
and a slight softening to $\Gamma=2.78^{+0.16}_{-0.11}$ above $E_{\mathrm{break}}$.
The spectrum shows that the peak in the SED was below $\sim3\,\mathrm{keV}$ during the observations.

Possible systematic errors of the PCU spectrum were investigated by comparing recent archived spectra from Cassiopeia A with data from
previous years, see the recommendations in the RXTE frequently asked 
questions\footnote{\em http://heasarc.gsfc.nasa.gov/docs/xte/ftools/xtefaq.html}.
For the purpose of the analysis presented here, we concluded that systematic errors are of no concern.

For the construction of the simultaneous SED discussed in Section \ref{S:discussion}, 
the X-ray spectrum derived from the entire RXTE data set was used, 
and similarly the H.E.S.S. spectrum from the entire March 2005 data set.
A restriction of both the H.E.S.S. and RXTE data to the strictly simultaneous times appears overcautious, given the steady
measured fluxes, and would have reduced the statistical accuracy,
since only
13.3 (54\%) of 24.9 hours RXTE observations are overlapping with H.E.S.S. data. These 13.3 hours conversely represent
37\% of the total March 2005 H.E.S.S. data set (35.6 hours of on-source observation time).
We note that the March 2005 data set presents -- in terms of simultaneity and statistical accuracy 
-- the best VHE/X-ray data set for 1ES\,1101-232 so far.

\begin{table*} 
\begin{center}
  \begin{tabular}{l|l|l|l|l} 
\hline
Satellite  & Observation ID & Observation Dates & Pointings     & Used Detectors And Live Time \\ \hline \hline
RXTE       & 91123          & March 5-16, 2005  & 48            & PCU\,2: 89.60\,ksec \\ \hline
XMM-Newton & 205920601      & June 8, 2004      & 1             & MOS\,2 (thin filter): 17.65\,ksec \\
           &                &                   &               & PN (thin filter): 17.01\,ksec \\
           &                &                   &               & OM: V 1.60, B 1.88, U 1.60, UVW1 4.00, UVM2 4.00\,ksec \\
\hline
  \end{tabular}
\end{center}
\caption{Observation log of the two X-ray satellite observations on 1ES\,1101-232 used for this paper.}
  \label{T:1101xraylog}
\end{table*}

\subsubsection{Optical data}
\label{SSS:rotse}

During the March 2005 campaign, 
optical data on 1ES\,1101-232 were obtained using the ROTSE 3c telescope \citep{akerlof2003rotse},
which is located at the H.E.S.S. site. 
The ROTSE 3c telescope is a fast slewing robotic telescope with a 45\,cm mirror and a wide field of view of $1.85^{\circ}$. 
To collect as much light as possible\footnote{
The main
purpose of the ROTSE telescopes are optical afterglow observations of gamma ray bursts.}, 
no optical filter is used; the ROTSE passband is 400 to 
900\,nm.

During each of the 11 nights, typically 18 frames of 60 seconds exposure time were obtained.
After quality selection, 141 frames were used for analysis.
The standard ROTSE calibration chain\footnote{\em http://rotse1.physics.lsa.umich.edu/equipment/} delivered bias-corrected, flatfielded images. 
To obtain object magnitudes, the standard ROTSE procedure involves a vignetting correction and a 
relative photometry to a large number of comparison objects in the field of view, using their USNO-A2 \citep{monet1998usnoa2} R-band
magnitudes as reference.
The standard ROTSE photometry however failed to analyze the data from 1ES\,1101-232 because of source confusion,
therefore a manual photometry was performed, as described in the following. 

Three temporally-stable comparison stars close to 1ES\,1101-232 were selected.
Two of these stars have been identified by
\citet{smith1991opticalcalibrators} as calibrators and have known photometric multi-color data,
the third star was used for additional cross-checks.
An aperture of $R=7.5''$ was used for photometry.
The stability of our photometry procedure was verified 
with several individual frames, by checking the correlation between measured fluxes and 
C-band magnitudes of the reference and various other field stars. C-band magnitudes ($579\textnormal{--}642\,\mathrm{nm}$ bandpass) were derived from the
CCD astrograph catalog UCAC2 \citep{zacharias2004ucac2}.

To obtain a light curve and check for variability of the optical flux,
for each frame an effective C-band magnitude of 1ES\,1101-232 was derived, using the three comparison stars as calibrators.
The optical light curve so derived exhibited only marginal flux variations, nightly flux averages showed changes below 0.1\,mag.

To derive an average optical flux of the BL Lac for use in the SED, an absolute flux calibration and subtraction of the
host galaxy flux had to be performed. However, two facts caused this to be very difficult: the wide ROTSE bandpass, and the measured
flux of $m_{\mathrm{C}}=16.03\,\mathrm{mag}$ which is close to the flux from the host galaxy ($m_{\mathrm{R}}=16.41\,\mathrm{mag}$).
Therefore, 
only an upper limit and a tentative lower limit of the optical flux could be derived.

In order to subtract the flux from the host galaxy, 
it was
verified that the spectra of the two comparison stars used
\citep{smith1991opticalcalibrators} are similar to template spectra of elliptical hosts \citep{fukugita1995colors}
at the redshift of 1ES\,1101-232. Applying the de Vaucouleurs profile with an effective radius of the host galaxy
of $R_{\mathrm{e}}=4.1''$ 
and its total flux of $m_{\mathrm{R}}=16.41\,\mathrm{mag}$ as measured by \citet{falomo2000optical},
we concluded that $\sim$40\% of the measured intensity comes from the host galaxy. 
If the BL Lac had also
a similar spectrum, the apparent magnitude of the BL Lac would then be $m_{\mathrm{R}}=16.4\,\mathrm{mag}$. 

However, the
wide bandpass of the ROTSE instrument causes this estimate to be quite susceptible to the actual BL Lac spectrum,
which is expected to be much
harder than the spectrum of the host galaxy.
To estimate the magnitude of this effect, 
power-law spectra
$S_{\nu}\propto\nu^{-\alpha}$ with 
$\alpha$ between 1.0 and -0.5
were simulated,
under the assumption of a flat or symmetric response 
of the ROTSE detector between 400\,nm and 900\,nm,
and 
correction factors between 1.20 and 1.44 to the R-band flux
were derived.

Magnitudes were finally corrected for Galactic extinction, using 
a B-band extinction $A_B=0.254\,\mathrm{mag}$ (provided by NED; from \citet{schlegel1998extinction})
and following the extinction laws of \citet{cardelli1989extinction}.
Fluxes were derived from the magnitudes using the absolute calibrations by \citet{bessel1979magnitudes}.
With this, an upper limit (assuming $\alpha=-0.5$) of $S_{\mathrm{R}}=2.6\,\mathrm{mJy}$ was derived,
and a tentative lower limit (using no correction factor) of $S_{\mathrm{R}}=1.0\,\mathrm{mJy}$.

\subsection{Observations in June 2004}
\label{SS:xmmobs}

The XMM-Newton observations on 1ES\,1101-232 were conducted on June 8, 2004, as part of the pointings for XMM-Newton proposal 20592. 
A continuous pointing of 19.6\,ksec was scheduled such that H.E.S.S. observations could be conducted simultaneously.
Simultaneous ROTSE 3c observations could not be performed because the telescope was under repair during that period.

\subsubsection{X-ray data}
\label{SSS:xmmxray}

The XMM-Newton data files were processed with \verb+xmmsas 6.5.0+ following standard 
procedures as described in the SAS Handbook and calibration documents \citep{kirsch2006xmm},
where also systematic errors are discussed.
For our analysis,
the most recent calibration files as available in January 2006 were used.
Spectral and timing analysis was performed with \verb+XSPEC 11.3.1d+
and \verb+XRONOS 5.21+, respectively.

The EPIC instruments during this pointing were set in timing 
(PN and MOS\,1 cameras) and small window (MOS\,2 camera) mode, since one of the goals was to
study spectral variability at the shortest possible timescales without pile-up problems,
especially if the source were found in a very bright state.

During this observation, however, the source was characterized by a constant flux
on all timescales. Analysis of the light curves extracted in different energy bands, and the 
corresponding hardness ratios, showed no indication of spectral variability as well.
Therefore 
the entire dataset 
was used
to derive a time-integrated energy spectrum. 
In the following, since the MOS\,1 data are in agreement with the other instruments
but present a higher noise, 
the analysis 
was restricted 
to PN and MOS\,2.

Pile-up effects in the PN and MOS\,2 data were checked with \verb+epatplot+, 
and a mild pile-up was found for MOS\,2, that can be removed considering
single pixel events only (\verb+PATTERN=0+), which were compared to spectra extracted
with patterns 0-12 from different annular source 
regions \citep{molendi2003epic}.
For the PN, since in timing mode, \verb+PATTERN+$\le$\verb+4+ was selected
(single and double pixel events). For both instruments, quality \verb+FLAG=0+ was used.
The total livetime after screening for soft proton flares was 17.005\,ksec for PN 
and 16.653\,ksec for MOS\,2.

For MOS\,2, source counts were extracted from a circle with radius 45'' 
centered on the source centroid (obtained through \verb+eregionanalyse+). 
The background was extracted from the nearest source-free region on 
the peripheral CCDs, with radius 90''. 
A check using different locations on the peripheral CCDs,
and the same source position on the blank-sky fields 
(available at the XMM-Newton Science Operating Center),
showed no relevant differences, as was expected given the source rate
($\sim8\,\mathrm{cts\,s^{-1}}$) and spectrum.
For the PN (which in timing mode has only one-dimensional imaging capabilities),
photons were selected from rows $27 \le \mathrm{RAWX} \le 47$ 
(i.e., $\pm10$ RAW pixels around the source strip),
and $2 \le \mathrm{RAWX} \le 18$ for the background.
To avoid the increased noise at low energies, the energy range for PN was restricted to 
$0.5\textnormal{--}10\,\mathrm{keV}$, while events down to $0.2\,\mathrm{keV}$ were accepted for MOS\,2,
as allowed by the latest calibration \citep{kirsch2006xmm}.
Spectra were rebinned with \verb+grppha+ to have at least 20 counts per channel. 
Response matrices and ancilliary files were produced with \verb+rmfgen+ and \verb+arfgen+.

The spectra were fitted with source models plus interstellar medium absorption,
using the model \verb+phabs+ with abundances from \citet{wilms2000abundances}.
The absorbing column density was fixed to the Galactic value 
$N_{\mathrm{H}}=5.76\times10^{20}\mathrm{cm^{-2}}$, 
but 
also 
a model with
additional free absorption at the source (z=0.186)
was tested.
The host galaxy of 1ES 1101-232 is not a normal elliptical galaxy, the measured boxy isophotes 
\citep{falomo2000optical} may be indicative of extra absorbers, though in previous 
X-ray observations especially with BeppoSAX such possibility was statistically disfavoured
(see Section \ref{SS:previousobs}). 

The PN and MOS\,2 spectra 
were inspected 
separately, 
but finally the data from the two instruments 
were fitted together, 
with a free constant to allow for a different 
normalization between the two instruments (which remained within 3\%).
The results from the combined fit are reported in Tab.\,\ref{T:1101xray}.
The $\chi^2$-values of the combined fits are relatively high, but to a large extent this is owing to
residual cross-calibration uncertainties (though strongly improved with SAS 6.5) combined with large statistics.

With Galactic absorption, a single power law does not provide an acceptable fit for the single detectors
as well as for the combined fit ($\chi^{2}_{\mathrm{red}}= 1.880$ for $1547\,\mathrm{d.o.f.}$).
A broken power-law model significantly improves all fits, with 
$\chi^{2}_{\mathrm{red}}= 1.195$ (1545 d.o.f., null-hypothesis probability $2 \cdot 10^{-7}$) for the combined fit. The hard 
($\Gamma_{1}=1.94\pm0.01$) and soft ($\Gamma_{2}=2.19\pm0.01$) spectral indices
locate the X-ray peak of the SED at the break energy $E_{\mathrm{break}}=1.1\,\mathrm{keV}$.
Looking into the individual camera fits, the soft slopes are in agreement
($\Gamma_{2,\mathrm{PN}}=2.19^{+0.01}_{-0.02}$ vs. $\Gamma_{2,\mathrm{MOS\,2}}=2.21^{+0.01}_{-0.02}$),
while break positions and hard slopes
are slightly differing ($\Gamma_{1,\mathrm{PN}}=1.98^{+0.01}_{-0.02}$, $E_{\mathrm{break,PN}}=1.09^{+0.01}_{-0.02}\,\mathrm{keV}$,
vs. $\Gamma_{1,\mathrm{MOS\,2}}=1.91^{+0.01}_{-0.02}$, $E_{\mathrm{break,MOS\,2}}=1.45^{+0.01}_{-0.02}\,\mathrm{keV}$);
the respective fits are well acceptable for PN but slightly disfavoured for MOS\,2 (null-hypothesis probability: PN 5\%, MOS\,2 0.3\%).
Reasons for the discrepancies are the different fit ranges towards the soft end of the spectra, and the 
already mentioned 
calibration uncertainties.

The combined broken power-law fit, considering $N_{\mathrm{H}}$ as a free parameter, 
yields a column density of $N_{\mathrm{H}} = 5.8\pm0.7\times10^{20}\mathrm{cm^{-2}}$
which is in very good agreement with the Galactic value. 
To test whether additional absorption at the source is compatible with the data, 
a single power law model with Galactic absorption and free absorption at z=0.186
was investigated.
The results are less favoured than the fits with pure Galactic absorption
(null-hypothesis probability: PN: 0.7\%, MOS\,2: $10^{-15}$, combined fit: $3 \cdot 10^{-11}$, extra absorption 
$N_{\mathrm{H}} \sim 3\times10^{20}\mathrm{cm^{-2}}$),
and provide more skewed residuals with an excess at lower energies less compatible with the hypothesis of higher absorption.
The difference 
to the models with Galactic absorption
is however not large, also considering the fact that
the MOS\,2 spectrum still exhibits some unmodeled residua at low energies ($0.4\textnormal{--}0.5\,\mathrm{keV}$), 
which could possibly be because of the mentioned detector calibration uncertainties and/or imperfect modeling of the hydrogen 
absorption
(even free abundances improve only slightly the residuals).

To obtain the unfolded energy spectrum for
the SED, 
the results of the combined fit were used.
The residuals of the MOS\,2 data to the respective MOS\,2 model were multiplied with the
model derived for the PN. With this prescription, the absolute flux calibration from PN is trusted, while the full spectral energy range 
from MOS\,2 can be used.
Finally, the data were rebinned logarithmically in energy. 
The unfolded spectrum is shown in Fig.\,\ref{F:1101xtespectra}, together with a similarly derived spectrum under the assumption of a
pure power law and additional absorption at the source, for comparison.
The model fits in Fig.\,\ref{F:1101xtespectra} were derived from the unfolded
spectra and are shown to indicate the size of the residuals.
We note that the unfolded spectra do not strongly depend on the exact shape of the modeled spectrum that is used in the unfolding
procedure (using for example a pure power law instead of a broken power law yields a compatible spectrum within errors).

To conclude, while small amounts of extra absorption at the source -- which would soften the slope
below 1\,keV -- cannot be excluded based on the XMM-Newton spectra alone, there is
good evidence that the X-ray spectrum from 1ES\,1101-232 exhibited a peak in the SED at $\sim$1\,keV during our observations, similar to
earlier X-ray observations of this source.

For the June 2004 SED discussed in the next section,
the data of the entire H.E.S.S. June data set (i.e., $\pm 3$ days from the XMM-Newton pointing) were taken,
in order to obtain a significant signal from the VHE data.
Quantitatively,
3.4 (66\%) of the 5.1 hours (MOS\,2) XMM-Newton observations have simultaneous H.E.S.S. data. 
These 3.4 hours conversely represent 37\% of the June 2004 H.E.S.S. data set (9.1 hours of on-source observation
time).

\subsubsection{Optical data}
\label{SSS:xmmoptical}

During the observations on 1ES\,1101-232, the optical monitor (OM)
onboard XMM-Newton took five exposures with five different filters, from the V to the UV band,
with a window in fast mode on our target.
Details are given in Tab.\,\ref{T:1101xraylog}. 
As in the EPIC instruments, no variability was found in any OM exposure.
Therefore,
a photometric spectrum 
was extracted 
from all filters.

OM data were processed with \verb+xmmsas 6.5.0+. 
Only 
data from the imaging mode 
were used 
in the following.
OM count rates were extracted using the point source analysis procedure of the OM photometry program.
An aperture of $R=6''$ 
was used
for the source, background counts were extracted from an annulus with $7''<R<12.5''$, for all filters.
Count rates were converted into
fluxes according to the prescriptions of the XMM-Newton watch-out pages\footnote{\em http://xmm.vilspa.esa.es/sas/new/watchout}.
The conversion
factors for a white dwarf
were used,
as recommended by the OM calibration scientist (Nora Loiseau, priv. comm.). 

The point spread function of the OM is
considerably wider in the UV (UVW1 and UVM2 filters) than in the other filters, 
requiring specific aperture corrections in those bands.
The default aperture for these filters could not be used because of bright sources
in the corresponding background annulus, therefore 
the same source and background apertures 
were used
as for the other filters.
The validity of the aperture correction was tested by selecting different source and background aperture sizes.
The systematic error on fluxes derived with the UV filters 
was estimated
to be $\sim$20\%. 

In the V and B bands, a contribution from the host galaxy needs to be corrected. Using the same de Vaucouleurs profile as
in Section \ref{SSS:rotse}, 
61\% and 15\% of the host galaxy flux 
were estimated
to fall into the source and background apertures, respectively.
These fractions of 
the host galaxy flux 
were subtracted,
using its R-band magnitude $m_{\mathrm{R}}=16.41\,\mathrm{mag}$ from \citet{falomo2000optical} and 
the elliptical galaxy spectral template from \citet{fukugita1995colors}.

Fluxes were finally corrected for Galactic extinction, again using $A_B=0.254\,\mathrm{mag}$.
The derived optical spectrum is shown in the SED in Fig.\,\ref{F:1101sed}, lower left panel. 
Error bars at the two UV flux points are from our systematic uncertainty estimate.
We note that the U-filter flux is probably the most reliable flux estimate for the BL Lac, because of the mentioned uncertainties
in the UV filters and because of the host galaxy influence at larger wavelengths.

\begin{table*} 
\begin{center}
  \begin{tabular}{l|l|l|ll|lll|l} 
\hline
Obs.                  & $F_{2-10\,\mathrm{keV}}~\times$                                 & $\Gamma_{N_{\mathrm{H,gal.}}}$           & $\Gamma_{N_{\mathrm{H,free}}}$          & $N_{\mathrm{H}}~\times $                        & $\Gamma_1$                 & $E_{\mathrm{break}} $                & $\Gamma_2$                 & Reference \\
                      & $ 10^{-11}\mathrm{erg}\,\mathrm{cm}^{-2}\mathrm{s}^{-1}$        &                                          &                                         &                $10^{20}\mathrm{cm}^{-2}$        &                            & $\mathrm{keV}$                       &                            &           \\ \hline
ROSAT                 & $2.33^{+0.30}_{-0.27}$                                          & $2.23^{+0.04}_{-0.03}$ (a)               & $2.43^{+0.08}_{-0.08}$                  & $6.8^{+0.3}_{-0.4}$                             & $2.07^{+0.09}_{-0.22}$     & $0.69^{+0.24}_{-0.25}$               & $2.42^{+0.12}_{-0.09}$     & (1) \\
SAX I                 & $3.76$                                                          & $1.97^{+0.03}_{-0.02}$ (b)               & $2.03^{+0.05}_{-0.04}$                  & $8.9^{+3.7}_{-2.7}$                             & $1.59^{+0.15}_{-0.14}$     & $1.36^{+0.29}_{-0.25}$               & $2.05^{+0.03}_{-0.04}$     & (2) \\
SAX I$^\dagger$       & $3.68$                                                          & -                                        & $2.01^{+0.06}_{-0.06}$                  & $8.5^{+3.2}_{-2.0}$ $^{\dagger\dagger}$         & $1.73^{+0.13}_{-0.09}$     & $3.45^{+0.66}_{-0.52}$               & $2.19^{+0.14}_{-0.11}$     & (3) \\
SAX II                & $2.55$                                                          & $2.19^{+0.03}_{-0.03}$ (c)               & $2.25^{+0.04}_{-0.03}$                  & $8.3^{+2.1}_{-1.2}$                             & $1.80^{+0.16}_{-0.22}$     & $1.34^{+0.48}_{-0.26}$               & $2.29^{+0.04}_{-0.04}$     & (2) \\
XMM I                 & $2.3$                                                           & $2.23^{+0.02}_{-0.01}$ (d)               & $2.40^{+0.02}_{-0.03}$                  & $10.9^{+0.6}_{-0.6}$                            & $^{*}2.06^{+0.12}_{-0.12}$ & -                                    & $^{*}2.38^{+0.16}_{-0.15}$ & (4) \\
XMM II                & $3.74^{+0.01}_{-0.02}$                                          & $2.11^{+0.01}_{-0.01}$ (e)               & $2.19^{+0.01}_{-0.01}$                  & $8.53^{+0.17}_{-0.17}$                          & $1.94^{+0.01}_{-0.01}$     & $1.11^{+0.05}_{-0.04}$               & $2.19^{+0.01}_{-0.01}$     & this work \\ \hline
\hline
XTE                   & $5.07\pm 0.02$                                                  & -                                        & -                                       & -                                               & $2.49^{+0.02}_{-0.02}$     & $7.9^{+1.0}_{-0.9}$                  & $2.78^{+0.16}_{-0.11}$     & this work \\
\hline
  \end{tabular}
\end{center}
\caption{Spectral fit results of selected X-ray observations on 1ES\,1101-232. (a-e): Values from spectral fits with low probablility: 
(a):0.7\%. (b):$<$2\%. (c):$<$0.5\%. (d):$<$$10^{-9}$ (e):$<<$$10^{-10}$.
$^\dagger$: Reanalysis of the data set ``SAX I'' presented originally by \citet{wolter20001101}. 
$^{\dagger\dagger}$: \citet{donato2005saxcatalogue} used $N_{\mathrm{H,gal.}}=6.05\times 10^{20}\mathrm{cm}^{-2}$ and additional absorption at the source ($2.4^{+3.2}_{-2.0}\times 10^{20}\mathrm{cm}^{-2}$);
for simplicity,
the sum of these two values 
is quoted 
here.
$\Gamma_1$, $E_{\mathrm{break}}$, $\Gamma_2$ from a broken power-law fit, except
$^{*}: \Gamma_{1}=\Gamma_{0.5\textnormal{--}1.0\,\mathrm{keV}}, \Gamma_{2}=\Gamma_{4.0\textnormal{--}10.0\,\mathrm{keV}}$.
The XMM II values are from the combined PN/MOS\,2 fits as described in the text, the flux normalisation is from PN only.
The XTE row is detached to emphasize that the data are not sensitive to the break seen by all other instruments 
in the $\sim 0.5\textnormal{--}3.5\,\mathrm{keV}$ range.
References: 
(1) \citealt{wolter1998bepposax};     
(2) \citealt{wolter20001101};         
(3) \citealt{donato2005saxcatalogue}; 
(4) \citealt{perlman2005curvature}.   
}
  \label{T:1101xray}
\end{table*}

\subsection{Previous X-ray observations of 1ES\,1101-232}
\label{SS:previousobs}

1ES\,1101-232 has already been observed in previous pointed X-ray observations with ROSAT \citep{wolter1998bepposax},
BeppoSAX \citep{wolter20001101} and XMM-Newton
\citep{perlman2005curvature}. Spectra and fluxes were significantly different comparing individual observations, but the
flux changes were not large ($\pm 25$\%). 
For example, the two spectra taken with BeppoSAX in 1997 and 1998 (dubbed as high and low state in \citet{wolter20001101})
only showed a significant change above the peak at $\sim1.4\,\mathrm{keV}$, with a total flux difference of 50\%.
Results from the different observations, including the two observations discussed in this paper, are summarized in Tab.\,\ref{T:1101xray}.

In all observations, pure intrinsic power laws were rejected if only Galactic absorption values were assumed. 
All observations could
be modeled with a pure power law, under the assumption of extra absorption of the order of 
$N_{\mathrm{H}} \simeq 3 \times 10^{20}\mathrm{cm}^{-2}$. 
Nevertheless, in cases where the statistical accuracy was sufficient (SAX\,I, SAX\,II, XMM\,II, cf. Tab.\,\ref{T:1101xray}), 
significantly better fits were derived, 
using only Galactic absorption and a curved intrinsic spectrum (usually modeled by a broken power law).

Assuming hence no extra absorption at the source,
the low-energy peak in the SED of 1ES\,1101-232 has in most observations been located in the $\sim
0.5-3.5\,\mathrm{keV}$ range. 
From the difference of the position of the spectral break derived from the data set SAX\,I (see
Tab.\,\ref{T:1101xray}) by \citet{wolter20001101} on the one hand and by \citet{donato2005saxcatalogue} on the other hand, 
and also following the discussion in \citet{perlman2005curvature}, 
we conclude that a broken power law is only an approximation of the real spectrum. 
Nevertheless, this does not affect the SED modeling presented in the next section, 
as the used unfolded XMM-Newton X-ray spectrum 
does not significantly change if the broken power law model is replaced by a
curved model, such as the one described in \citet{fossati2000spectra421}.

\section{Discussion}
\label{S:discussion}

\subsection{Constraints on the extragalactic background light}
\label{SS:ebl}

The detection of VHE $\gamma$-ray emission from 1ES\,1101-232 
was used to constrain the
density of the EBL flux in the 
wavelength range of $\sim 0.75\textnormal{--}3.5\,\mathrm{\mu m}$
(\citeauthor*{aharonian2006ebl} and Section \ref{SS:vheebllimit}).
The
measured average photon index 
of $\Gamma=2.9$
either precludes high EBL levels, or indicates a much harder instrinsic spectrum
than seen or expected in other VHE blazars. 
For the purpose of this paper, 
the idea 
was adopted
that the intrinsic VHE $\gamma$-ray spectrum of 1ES\,1101-232 is restricted
to $\Gamma_{\mathrm{VHE,deabs}} \ge 1.5$.
This limit is empirically derived in all blazars of lower redshift 
(where EBL corrections are less severe).
It is also theoretically expected in present standard leptonic
or hadronic scenarios for blazar emission employing shock acceleration models \citep{malkov2001acctheory},
because the hardest energy index obtained for accelerated particles is $p=1.5$ which results in $\Gamma \ge 1.5$ for
all emission processes which can realistically be assumed (\citeauthor*{aharonian2006ebl}).
As shown in \citeauthor*{aharonian2006ebl}, the limit of $\Gamma_{\mathrm{VHE,deabs}} \ge 1.5$
applied to the measured spectrum of 1ES\,1101-232 results in an EBL upper limit of 
$\nu F_{\nu}=14\,\mathrm{nWm^{-2}sr^{-1}}$ at $1.5\,\mathrm{\mu m}$ for typical EBL shapes.

We note that harder photon spectra result if the low energy end of the particle distribution is altered from the assumptions
mentioned above. A variety of possibilities have been discussed in the literature.
Relativistic Maxwellian (pileup) particle energy spectra, produced in a balance of stochastic acceleration and radiative cooling
\citep{schlickeiser1985pileup,henri1991pileupagn}, have previously been
invoked in the context of blazar spectra \citep{sauge2004pileup501}.
Radiative cooling of an injection spectrum flatter than $\gamma^{-2}$ could also result in a pileup spectrum
\citep{pinkau1980ngc4151,kardashev1962nonthermalspectra}.
Also bulk motion Comptonization of ambient low-frequency photons by a 
cold unshocked ultrarelativistic jet with a very large Lorentz factor of the order of
$10^{6}..10^{7}$ could yield a narrow, line-like spectrum \citep{aharonian2002mrk501}.

\citet{katarzynski2006ebl} used the total VHE $\gamma$-spectrum of 1ES\,1101-232 from \citeauthor*{aharonian2006ebl},
together with archival X-ray data that were not taken simultaneously with the H.E.S.S. data, 
and showed that emission from a power-law type particle spectrum with a low energy cut-off at large Lorentz factors $\sim 10^{5}$
can produce an intrinsic VHE $\gamma$-spectrum as hard as $\propto \nu^{+\frac{1}{3}}$.
This would allow for an EBL level of $\nu F_{\nu}(1.5\,\mathrm{\mu m})\simeq20\,\mathrm{nWm^{-2}sr^{-1}}$ and fit the VHE data.
We note that our simultaneous data show no indication for such effects in the synchrotron branch. For the purpose of this paper,
we refrain from discussing such spectra further.

\subsection{Spectral energy distribution of 1ES\,1101-232}
\label{SS:seds}

For the construction of the SED, 
the deabsorbed data
were taken,
using what we consider the best available deabsorption prescription.
The optical and X-ray data were corrected for Galactic absorption, 
see Sections \ref{SSS:rotse}, \ref{SSS:xmmxray}, and \ref{SSS:xmmoptical}.
The VHE $\gamma$-ray spectra were deabsorbed, 
using two possible levels of the present day EBL photon field: 
the {\em maximum}
value of
$\nu F_{\nu}=14\,\mathrm{nWm^{-2}sr^{-1}}$ at $1.5\,\mathrm{\mu m}$,
and a minimum value of $10\,\mathrm{nWm^{-2}sr^{-1}}$ corresponding to the lower limit placed by Galaxy counts
\citep{madau2000galaxycounts}.
We note that the galaxy counts are presumably below the actual EBL density because of missed light \citep{totani2001ebl}.
To derive the optical depth for the VHE $\gamma$-rays, the 
phenomenological EBL curve as used in \citeauthor*{aharonian2006ebl} was applied, 
after scaling to match the given maximum and minimum EBL densities and
to take galaxy evolution effects into account, as explained in Section \ref{SS:vhespectrum}.
A redshift correction of the frequencies shown in the SED 
(to account for the difference of apparent and restframe wavelength) was not performed,
as the correction would be dominated by the uncertainty of the emitting region's Doppler factor.

\begin{figure*} 
\centering
  \includegraphics[width=17cm]{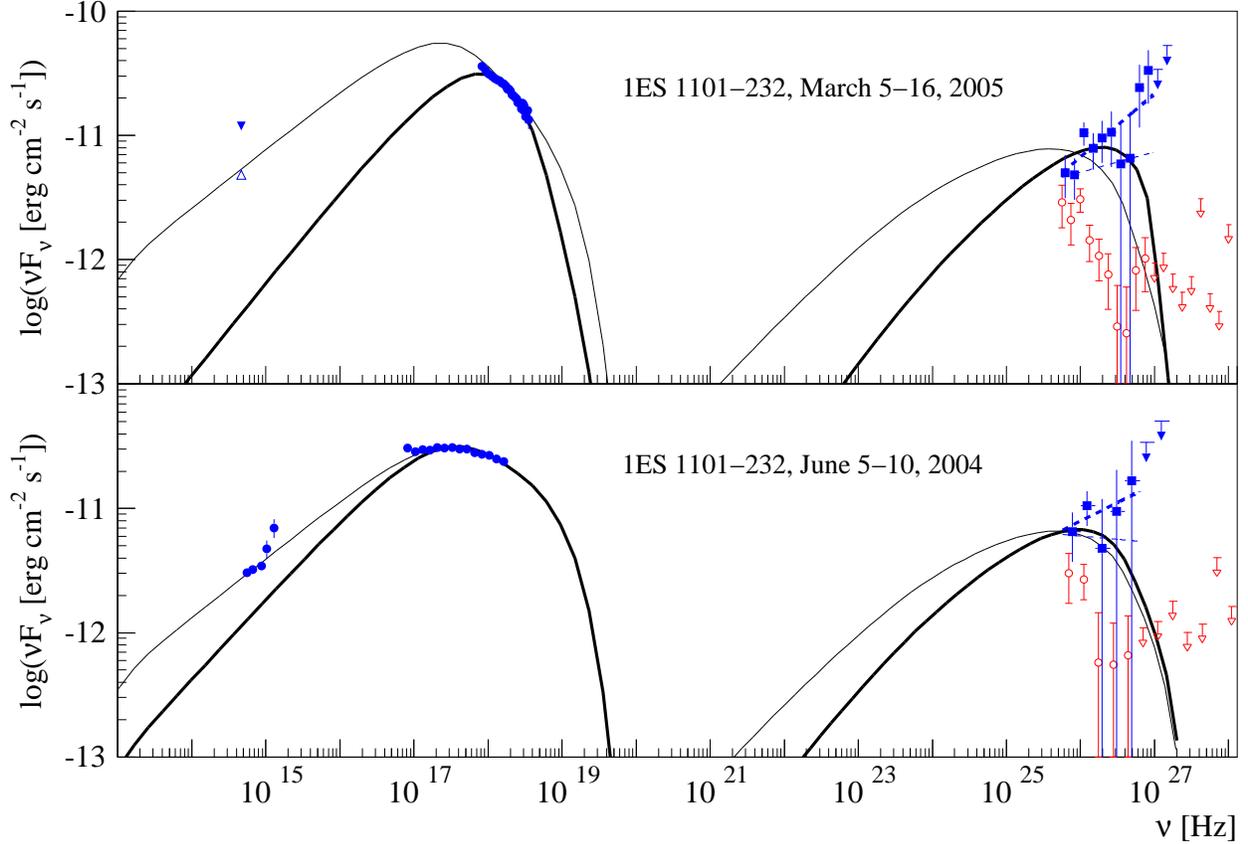}
\caption{
Spectral energy distribution of 1ES\,1101-232. {\bf Upper panel:} Data from March 5-16, 2005. 
X-ray data are from RXTE. In the optical band, an upper limit (filled triangle) and a tentative lower limit (open triangle)
from ROTSE 3c data are shown, see Section \ref{SSS:rotse} for details. 
In the VHE band, the measured H.E.S.S. spectrum (red, open symbols) and the deabsorbed spectrum
using a maximum EBL level of $14\,\mathrm{nWm^{-2}sr^{-1}}$ at $1.5\,\mathrm{\mu m}$ (see text) are shown;
for better visibility, all open symbols were slightly shifted to the left, to 90\% of the respective original frequency.
The thick dashed line is a power-law fit to the deabsorbed data as plotted,
while the thin dashed 
line indicates the effect if the EBL is lowered to the minimum level of $10\,\mathrm{nWm^{-2}sr^{-1}}$.
The latter value corresponds to the minimum level of the EBL as derived from known resolved galaxy counts.
Thick and thin solid curves
denote results from a single zone SSC model. The thick curves represent a model that was optimized to fit the
H.E.S.S. and X-ray data, while the thin lines denote a model with an electron distribution $N_{\gamma} \propto \gamma^{-2}$ below the break.
{\bf Lower panel:} Data from June 2004. X-ray and optical data were derived from an XMM-Newton
pointing on June 8, 
2004.
In the VHE band, 
H.E.S.S. data taken between June 5-10, 
2004,
are shown, using the same procedure as described for the upper panel.
Solid curves denote results from a single zone emission model, also following the same prescription as for the upper panel.
}
  \label{F:1101sed}
\end{figure*}

In Fig.\,\ref{F:1101sed} 
the SEDs of 1ES\,1101-232 
are shown for the two periods for which 
broadband data together with the H.E.S.S. VHE measurements
have been obtained. 
The upper panel of Fig.\,\ref{F:1101sed} contains the 
average fluxes of the March 2005 campaign (i.e., March 5-16, 2005), as derived from H.E.S.S., RXTE, and ROTSE 3c data.
As described in Section \ref{SSS:xtexray}, 
the data 
have not been restricted
to true simultaneity (i.e., on minutes timescale). However, all data were taken during the same observing nights, 
with no significant (VHE) or only mild (X-ray, optical) variations between days, and
there is no
sign of variability in these data on timescales shorter than a day.

The SED shown in the lower panel of Fig.\,\ref{F:1101sed} has a lesser degree of simultaneity, because it contains the average VHE $\gamma$-ray spectrum 
obtained from the H.E.S.S. June 2004 observations (i.e., June 5-10, 2004),
together with the X-ray and optical data as derived from the XMM-Newton observation performed on June 8 (see Section \ref{SSS:xmmxray}).

\subsection{Model-independent considerations}
\label{SS:interpretation}

Independent of specific emission scenarios, the broadband data from 1ES\,1101-232 presented here
show some remarkable features:

{\em VHE peak:} 
The hard intrinsic VHE $\gamma$-ray spectrum ($\Gamma_{\mathrm{deabs}}\lesssim2$) 
requires that the VHE $\gamma$-ray flux peak is located at $>3\,\mathrm{TeV}$ and
that the measured VHE photons do come from below that peak,
unless the lowest possible EBL level is assumed and errors on the VHE $\gamma$-ray spectrum are exploited to the limits.
The VHE $\gamma$-ray spectrum does not show hints of curvature. 
A steepening 
towards higher energies
would indicate the position of the SED high-energy peak at $\sim 3\,\mathrm{TeV}$ (either
due to a break in the particle distribution, or because of Klein Nishina effects), but 
the data are insufficient to claim such a steepening.

{\em Comparison 2004/2005:} 
The source was in a different state in March 2005
compared with the June 2004 period, judging from the X-ray 
data. 
While the X-ray flux and spectrum in June 2004 were comparable to previous X-ray measurements
(see Tab.\ref{T:1101xray}),
the X-ray spectrum in March 2005 as measured with RXTE 
shows a higher flux and a 
fairly soft spectrum, softer than any previously measured X-ray spectrum from this source.
The synchrotron peak (usually at $\sim 0.5\textnormal{--}3.5\,\mathrm{keV}$) 
cannot have shifted to 
higher energies during the March 2005 high state,
the XTE spectrum precludes a peak above $\sim 3\,\mathrm{keV}$.
Despite the different X-ray flux states in March 2005 and June 2004,
there is
no evidence for a change of the VHE $\gamma$-ray spectrum between the two periods;
nevertheless, statistical errors from the June 2004 data sets could allow  
for a factor of up to about two flux difference between the two periods.

{\em X-ray variability:} 
The X-ray light curve in March 2005 only exhibits mild variations. 
Under the assumption of $\gamma^{-2}$-type injection spectra, it seems
unlikely that the soft RXTE spectrum from that period is governed by emission from a cooled
particle spectrum. The spectrum could reflect the cut-off of the acceleration spectrum, but depending
on the assumed scenario, other options like softer injection spectra \citep[e.g.,][]{ghisellini2002lbl} are also possible.

{\em Leptonic emission spectrum:} 
One can compare X-ray and VHE $\gamma$-ray spectral indices under the assumption that electrons from the same energy band 
with $N_{\gamma} \propto \gamma^{-p}$ are emitting X-rays via synchrotron and VHE photons via IC processes.
The rather flat X-ray spectrum over nearly two decades in energy, as seen with XMM-Newton in 2004, 
constrains $p$ to $2.9<p<3.4$, and therefore requires that the VHE $\gamma$-ray emission is in the Thompson regime 
($1.9<\Gamma_{\mathrm{IC}}<2.2$) to be still in agreement with the measured VHE $\gamma$-ray spectrum 
($\Gamma_{\mathrm{Jun\,2004}}=1.70\pm0.47$). 
Assuming that the VHE $\gamma$-ray spectrum was constant throughout the observation periods ($\Gamma_{\mathrm{total}}=1.51\pm0.17$) would 
require either to drop the initial assumption about the common spectral range of the emitting electrons, 
or to lower the EBL to the lowest possible value of $\nu F_{\nu}(1.5\,\mathrm{\mu m})=10\,\mathrm{nWm^{-2}sr^{-1}}$.
For the March 2005 data set, 
a common
energy band of the 
electron spectrum cannot account for the X-ray and VHE $\gamma$-ray emission.

{\em Intrinsic source luminosity:} 
Depending on the assumed emission mechanism and EBL density, estimates of 
the intrinsic luminosity $L_{\mathrm{VHE}}$ of the VHE $\gamma$-ray emitting region in 1ES\,1101-232 can reach unusually large values.
Following \citet{sikora1997ssclumi}, 
the intrinsic luminosity would be $L_{\mathrm{VHE}} > 10^{39}\,\mathrm{erg\,s^{-1}}$ 
under the assumption of SSC VHE $\gamma$-ray emission from a moving blob with Doppler factor $\delta$$\sim$$30$.
Since however the emission seems
constant on timescales of days or maybe even years, a stationary jet model could also be invoked, leading to estimates of 
$L_{\mathrm{VHE}}>10^{42}\,\mathrm{erg\,s^{-1}}$. Adopting like in \citet{katarzynski2006ebl} the ``best fit" EBL model
from \citet{kneiske2004ebl}, 
with an EBL density approaching $\nu F_{\nu}(1.5\,\mathrm{\mu m}) \sim 20\,\mathrm{nWm^{-2}sr^{-1}}$,
would increase the intrinsic luminosity estimates to $> 10^{40}\,\mathrm{erg\,s^{-1}}$ for the moving blob and even
$> 10^{43}\,\mathrm{erg\,s^{-1}}$ for the stationary scenario.

\subsection{One-zone leptonic emission model}
\label{SS:model}

\begin{table} 
\begin{center}
  \begin{tabular}{l|l|l|l|l} 
\hline
                         & 2005 I                & 2005 II               & 2004 I                 & 2004 II               \\ \hline
$p_1$                    & 2                     & 1.5                   & 2                      & 1.7                   \\
$p_2$                    & 4.3                   & 7.0                   & 3.6                    & 3.6                   \\
$\gamma_{\mathrm{b}}$    & $1.8 \cdot 10^{5}$    & $3.8 \cdot 10^{5}$    & $1.8 \cdot 10^{5}$     & $1.8 \cdot 10^{5}$    \\
$K [\mathrm{cm^{-3}}]$   & $3.5 \cdot 10^{2}$    & $15$                  & $9.0 \cdot 10^{2}$     & $40$                  \\
$R [\mathrm{cm}]$        & $2.8 \cdot 10^{16}$   & $5.7 \cdot 10^{15}$   & $1.7 \cdot 10^{16}$    & $1.15 \cdot 10^{16}$   \\ \hline
$\delta$                 & \multicolumn{4}{|c}{25} \\
$B [\mathrm{G}]$         & \multicolumn{4}{|c}{$0.1$} \\
$\gamma_{\mathrm{min}}$  & \multicolumn{4}{|c}{$10^{3}$} \\
$\gamma_{\mathrm{max}}$  & \multicolumn{4}{|c}{$10^{6}$} \\
\hline
  \end{tabular}
\end{center}
\caption{Physical parameters of the SSC one-zone model spectra. 
2005 I and 2004 I, i.e. the scenarios with $p_{1}=2$, correspond to the models shown as thin solid lines in Fig.\,\ref{F:1101sed}.
2005 II and 2004 II correspond to the models shown as thick solid lines.
Doppler factor, magnetic field, $\gamma_{\mathrm{min}}$, and
$\gamma_{\mathrm{max}}$ are in all four cases identical
($H_{0}=75\,\mathrm{km\,s^{-1}Mpc^{-1}}$). 
}
  \label{T:1101sedparameters}
\end{table}

Neither flux correlations between the X-ray and VHE bands nor fast variability (i.e., on sub-day or even sub-hour timescale)
could be established with the present data set.
Therefore, there is no clear indication that would favour one-zone emission models, either leptonic
or hadronic.
Past experience
\citep[e.g.,][]{aharonian20052155mwl} has shown that both hadronic and leptonic scenarios have enough free parameters to generally
be able to explain
the broad-band emission from blazars, if no further arguments from variability can be invoked. Nevertheless, the reconstruced  
hard VHE $\gamma$-ray spectrum from 1ES\,1101-232 challenges one-zone emission models.

A simple leptonic emission model
was used, a time-independent SSC model as described by \citet{katarzynski2001ssc},
in an attempt to describe the measured VHE, X-ray, and optical data.
The description of a one-zone homogeneous, spherical emitting region $R$ and a homogeneous magnetic field $B$
was adopted, 
which propagates with Doppler factor
$\delta$ towards the observer. The high-energy electron distribution is modeled with a broken power law between Lorentz factors
$\gamma_{\mathrm{min}}$ and $\gamma_{\mathrm{max}}$ with a break at $\gamma_{\mathrm{b}}$, and a density normalisation $K$. 
The two epochs have been considered independently.
In the March 2005 data set, the most stringent constraint on the SED comes from the 
hard $\gamma$-ray spectrum extending towards 3\,TeV.
In the June 2004 case, the $\gamma$-ray part of the spectrum is easier to reproduce because of the smaller number of events and the 
slightly softer spectrum,
while the X-rays are more constraining with a rather flat spectrum over a large frequency range. 

The results from two parameter sets for each SED 
are shown
in Fig.\,\ref{F:1101sed} and Tab.\,\ref{T:1101sedparameters}.
In case I, $p_{1}$ (the energy index of the particles between $\gamma_{\mathrm{min}}$ and $\gamma_{\mathrm{b}}$) 
was set to 2, as expected from an uncooled, shock-accelerated particle distribution.
With $\gamma_{\mathrm{b,2005}}= 1.8 \cdot 10^{5}$ and $p_{2,2005} = 4.3$ ($p_{2}$ being the energy index between $\gamma_{\mathrm{b}}$
and $\gamma_{\mathrm{max}}$) for the March 2005 SED,
and $\gamma_{\mathrm{b,2004}} = 1.8 \cdot 10^{5}$, $p_{2,2004} = 3.6$ for the June 2004 data,
good fits 
were obtained
to the X-ray and optical data, respectively.
As expected, the hard VHE spectra are not well reproduced when using $p=2$.
The 2004 VHE data are still satisfactorily matched,  
but the very hard 2005 VHE $\gamma$-ray spectrum (reconstructed with the EBL density $14\,\mathrm{nWm^{-2}sr^{-1}}$),
having also smaller statistical errors, is not well reproduced by the model.
A simple $\chi^{2}$ test only yields a just over $2\,\sigma$ compatibility between the data
and the model. A lower EBL level 
improves the agreement with the data.

In case II, 
$p_{1}$ was chosen to be of the order of 1.5, which can be expected
for instance from particle acceleration at strong shocks in a relativistic gas.
Better fits are then obtained for the $\gamma$-ray spectra at both epochs as illustrated in Fig.\,\ref{F:1101sed},
especially for 2005 where the $\chi^{2}$ test now yields 
a $1\,\sigma$ 
compatibility. 
In this scenario, the optical flux 
cannot be described within the SSC one-zone model, and
has to be attributed to another component.
This additional low frequency emission could come
for example from some extended jet emission. 
Such an additional component is in any case needed to explain the emission at radio frequencies, 
which was measured with the Nancay radio telescope at $2.685\,\mathrm{GHz}$ 
(see \cite{aharonian20052155mwl} for a description of the instrument and data processing). The flux levels
obtained in June 2004 ($0.11\pm0.02\,\mathrm{Jy}$, not simultaneous to the other data presented in this paper) 
and in March 2005 ($0.08\pm0.01\,\mathrm{Jy}$, simultaneous to the data presented 
in this paper) were comparable to previous measurements
\citep{griffith1994parkessurvey,douglas1996texassurvey}.

To conclude, SSC one-zone models are globally able to reproduce the SED of 1ES\,1101-232 from the X-rays
to VHE $\gamma$-rays for the data set analyzed here.
For the 2005 data set, an inclusion of the optical lower limit, obtained through the ROTSE 3c detector, 
is however statistically disfavoured.
Moreover we should stress
that
with 1ES\,1101-232 the limit of the capabilities of SSC one-zone models
is reached,
as also discussed in \citeauthor*{aharonian2006ebl}.
It is
very difficult to get good fits for the shape of the
VHE $\gamma$-ray tail of the observed spectra, as long as one keeps usual assumptions for particle acceleration mechanisms
and does not assume an EBL level as low as $10\,\mathrm{nWm^{-2}sr^{-1}}$.
The generated spectra 
deviate, especially for the March 2005 data, from the hard spectra obtained with H.E.S.S.
Smaller statistical errors on the VHE $\gamma$-ray spectrum or an extension to higher energies (or both),
further constraints on the size of the emitting zone from variability data,
or 
a slight increase of the absorption by extragalactic background 
above the value of $14\,\mathrm{nWm^{-2}sr^{-1}}$,
may reach the limit of
one-zone SSC models.
More complex (e.g., two-zone) scenarios may therefore be required.

\section{Conclusion}
\label{S:conclusion}

Using the H.E.S.S. Cherenkov telescope system, 
VHE $\gamma$-ray emission from 1ES\,1101-232
was discovered.
1ES\,1101-232 is
currently
the farthest object (z=0.186) with confirmed redshift detected in VHE $\gamma$-rays.
The VHE $\gamma$-ray spectrum favours a low level of the extragalactic background light at a few microns,
close to the lower limit placed by galaxy counts. 

The VHE data from 1ES\,1101-232 were taken together with measurements in the X-ray and
optical bands. The best SED from truly simultaneous observations
could be constructed from a multifrequency campaign performed in March 5-16, 2005. 
The data
revealed that the source was brighter in X-rays during this campaign than in any other previous X-ray observation,
but did not show strong flaring activity in either the X-ray or
the VHE band. The 
H.E.S.S. data set 
is compatible with
constant VHE $\gamma$-ray emission throughout all observation periods.

Using an EBL density of $\nu F_{\nu}=14\,\mathrm{nWm^{-2}sr^{-1}}$ at $1.5\,\mathrm{\mu m}$ 
to deabsorb the VHE $\gamma$-ray spectrum, the H.E.S.S. data result in a very hard
intrinsic spectrum of 1ES\,1101-232, with a peak in the VHE power output above 3\,TeV.
The spectrum is harder ($\Gamma\simeq1.5$) than in 
parametrisations using 
SSC models.
An EBL level below $14\,\mathrm{nWm^{-2}sr^{-1}}$ would soften the $\gamma$-ray spectrum nearly to $\Gamma\simeq2.0$, 
which would ease the application of SSC scenarios but at the same time challenge EBL models.
On the other hand, an EBL density above $14\,\mathrm{nWm^{-2}sr^{-1}}$ would result in an even harder $\gamma$-ray spectrum, therefore 
such high EBL levels can be excluded under the assumption that 1ES\,1101-232 is similar to all VHE blazars known so far
(\citeauthor*{aharonian2006ebl}). 
A $\gamma$-ray spectrum in 1ES\,1101-232 harder than $\Gamma\simeq1.5$ would indicate previously unseen blazar physics, 
but the broadband data presented in this paper currently do not support such a conclusion.

In conclusion we find 1ES\,1101-232 to be at an extreme end of blazar 
characteristics. This is the first time a detected blazar spectrum peaks 
above the VHE band. The data challenge current blazar models in the
high-energy end of the electromagnetic spectrum.

\acknowledgements{
The support of the Namibian authorities and of the University of Namibia
in facilitating the construction and operation of H.E.S.S. is gratefully
acknowledged, as is the support by the German Ministry for Education and
Research (BMBF), the Max Planck Society, the French Ministry for Research,
the CNRS-IN2P3 and the Astroparticle Interdisciplinary Programme of the
CNRS, the U.K. Particle Physics and Astronomy Research Council (PPARC),
the IPNP of the Charles University, the South African Department of
Science and Technology and National Research Foundation, and by the
University of Namibia. We appreciate the excellent work of the technical
support staff in Berlin, Durham, Hamburg, Heidelberg, Palaiseau, Paris,
Saclay, and in Namibia in the construction and operation of the
equipment.

We thank the ROTSE collaboration for providing the ROTSE 3c optical data,
and L. Ostorero for help with the optical data analysis. 
This research has made use of the NASA/IPAC Extragalactic Database (NED).
The R-band image around 1ES\,1101-232, taken with the UK Schmidt telescope,
was retrieved from the digital sky survey provided by the ST ScI.
This work uses data obtained at the Nan\c{c}ay Radio Telescope.}

\bibliographystyle{aa.bst}

\begin{thebibliography}{4}
\bibitem[Abraham et al.(1991)]{abraham1991optical} Abraham, R. G., McHardy, I. M., \& Crawford, C. S., MNRAS, 1991, 252, 482
\bibitem[Aharonian(2000)]{aharonian2000protonjet} Aharonian, F. A., New Astronomy, 2000, 5, 377
\bibitem[Aharonian(2001)]{aharonian2001proceedings} Aharonian, F. A., {\em Invited, Rapporteur, and Highlight Papers, Proc. 27th ICRC (Hamburg)}, ed. R. Schlickeiser, 2001, p. 250, astro-ph/0112314
\bibitem[Aharonian et al.(1999)]{aharonian1999hegra501} Aharonian, F. A., et al. (The HEGRA collaboration), A\&A, 1999, 342, 69
\bibitem[Aharonian et al.(2002a)]{aharonian2002mrk501} Aharonian, F. A., Timokhin, A. N., \& Plyasheshnikov, A. V., A\&A, 2002a, 384, 834
\bibitem[Aharonian et al.(2002b)]{aharonian2002hegra421} Aharonian, F. A., et al. (The HEGRA collaboration), A\&A, 2002b, 393, 89
\bibitem[Aharonian et al.(2005a)]{aharonian20052155mwl} Aharonian, F. A., et al. (H.E.S.S. Collaboration), A\&A, 2005a, 442, 895
\bibitem[Aharonian et al.(2005b)]{aharonian2005standardanalysis} Aharonian, F. A., et al. (H.E.S.S. Collaboration), A\&A, 2005b, 430, 865
\bibitem[Aharonian et al.(2006a)AHA06a]{aharonian2006ebl} Aharonian, F. A., et al. (H.E.S.S. Collaboration), Nature, 2006a, 440, 1018 (AHA06a)
\bibitem[Aharonian et al.(2006b)]{aharonian20052356} Aharonian, F. A., et al. (H.E.S.S. Collaboration), A\&A, 2006b, 455, 461
\bibitem[Aharonian et al.(2006c)]{aharonian2006crab} Aharonian, F. A., et al. (H.E.S.S. Collaboration), A\&A, 2006c, 457, 899
\bibitem[Akerlof et al.(2003)]{akerlof2003rotse} Akerlof, C. W., Kehoe, R. L., McKay, T. A., et al., PASP, 2003, 115, 132
\bibitem[Albert et al.(2006)]{albert2006magic1959} Albert, J., Aliu, E., Anderhub, H. et al., ApJ., 2006, 639, 761
\bibitem[Band \& Grindlay(1986)]{band1986ecfields} Band, D. L., \& Grindlay, J. E., ApJ., 1986, 308, 576
\bibitem[Benbow(2005)]{benbow2005standardanalysis} Benbow, W., for the H.E.S.S. Collaboration, {\em Towards a Network of Atmospheric Cherenkov Detectors VII}, ed. B. Degrange \& G. Fontaine, 2005, p. 163
\bibitem[Bessel(1979)]{bessel1979magnitudes} Bessel, M. S., 1979, PASP, 91, 589
\bibitem[Biermann \& Strittmatter(1987)]{biermann1987protonjet} Biermann, P. L., \& Strittmatter, P. A., ApJ., 1987, 322, 643
\bibitem[Blackburn(1995)]{blackburn1995ftools} Blackburn, J. K., 1995, {\em ASP conference series, Vol. 77, eds. R. A. Shaw, H. E. Payne, and J. J. E. Hayes}
\bibitem[B{\l}a\.zejowski et al.(2005)]{blazejowski2005whipple421} B{\l}a\.zejowski, M., Blaylock, G, Bond, I. H., ApJ, 2005, 630, 130
\bibitem[Buckley et al.(1985)]{buckley1985opticalidentification} Buckley, D. A. H., Tuhoy, I. R., \& Remillard, R. A., Proc. ASA, 1985, 6, 147 
\bibitem[Cardelli et al.(1989)]{cardelli1989extinction} Cardelli, J. A., Clayton, G. C., \& Mathis, J. S., Apj., 1989, 345, 245
\bibitem[Chadwick et al.(1999)]{chadwick1999limits} Chadwick, P. M., Lyons, K., McComb, T. J. L., et al., ApJ, 1999, 521, 547
\bibitem[Coppi \& Aharonian(1999)]{coppi1999ebltev} Coppi, P. S., \& Aharonian, F. A., ApJ, 1999, 521, L33
\bibitem[Costamante \& Ghisellini(2002)]{costamante2002tevcandidate} Costamante, L., \& Ghisellini, G., A\&A, 2002, 384, 56 
\bibitem[Dermer \& Schlickeiser(1993)]{dermer1993ecblazarmodel} Dermer, C. D., \& Schlickeiser, R., ApJ., 1993, 416, 484
\bibitem[Donato et al.(2001)]{donato2001blazarsequence} Donato, D., Ghisellini, G., Tagliaferri, G., \& Fossati, G., A\&A, 2001, 375, 739 
\bibitem[Donato et al.(2005)]{donato2005saxcatalogue} Donato, D., Sambruna, R. M., \& Gliozzi, M., A\&A, 2005, 433, 1163
\bibitem[Douglas et al.(1996)]{douglas1996texassurvey} Douglas, J. N., Bash, F. N., Bozyan, F. A., et al., AJ, 1996, 111, 1945
\bibitem[Elvis et al.(1992)]{elvis1992einsteinsurvey} Elvis, M., Plummer, D., Schachter, J., \& Fabbiano, G., ApJS, 1992, 80, 257 
\bibitem[Falomo et al.(1993)]{falomo1993opticalir} Falomo, R., Bersanelli, M., Bouchet, P., \& Tanzi, E. G., AJ, 1993, 106, 11 
\bibitem[Falomo et al.(1994)]{falomo1994opticalspectra} Falomo, R., Scarpa, R., \& Bersanelli, M., ApJS, 1994, 93, 125
\bibitem[Falomo \& Ulrich(2000)]{falomo2000optical} Falomo, R., \& Ulrich, M.-H., A\&A, 2000, 357, 91
\bibitem[Fossati et al.(2000)]{fossati2000spectra421} Fossati, G., Celotti, A., Chiaberge, M., et al., ApJ, 2000, 541, 166 
\bibitem[Fukugita et al.(1995)]{fukugita1995colors} Fukugita, M., Shimasaku, K., \& Ichikawa, T., PASP, 1995, 107, 945
\bibitem[Ghisellini et al.(1985)]{ghisellinietal1985ssc} Ghisellini, G., Maraschi, L., \& Treves, A., A\&A, 1985, 146, 204 
\bibitem[Ghisellini et al.(2002)]{ghisellini2002lbl} Ghisellini, G., Celotti, A., \& Costamante, L., A\&A, 2002, 386, 833 
\bibitem[Griffith et al.(1994)]{griffith1994parkessurvey} Griffith, M. R., Wright, A. E., Burke, B. F., and Ekers, R. D., ApJS, 1994, 90, 179
\bibitem[Henri \& Pelletier(1991)]{henri1991pileupagn} Henri, G., \& Pelletier, G., ApJ, 1991, 383, L7
\bibitem[Hinton et al.(2004)]{hinton2004hess} Hinton, J. A., et al.  (H.E.S.S. Collaboration), New Astron. Rev., 2004, 48, 331 (astro-ph/0403052)
\bibitem[Hinton et al.(2005)]{hinton2005background} Hinton., J., Berge, D., \& Funk, S., and the H.E.S.S. Collaboration, {\em Towards a Network of Atmospheric Cherenkov Detectors VII}, ed. B. Degrange \& G. Fontaine, 2005, p. 183
\bibitem[Kardashev(1962)]{kardashev1962nonthermalspectra} Kardashev, N. S., Sov. Astron., 1962, 6, 317
\bibitem[Katarzy{\'n}ski et al.(2001)]{katarzynski2001ssc} Katarzy{\'n}ski, K., Sol, H., \& Kus, A., A\&A, 2001, 367, 809
\bibitem[Katarzy{\'n}ski et al.(2006)]{katarzynski2006ebl} Katarzy{\'n}ski, K., Ghisellini, G., Tavecchio, F., et al., MNRAS, 2006, 368, 52
\bibitem[Kirsch et al.(2006)]{kirsch2006xmm} Kirsch, M., at al., 2006, avaliable online at {\em http://xmm.vilspa.esa.es/docs/documents/CAL-TN-0018.pdf}
\bibitem[Kneiske et al.(2004)]{kneiske2004ebl} Kneiske, T. M., Bretz, T., Mannheim, K., Hartmann, D. H., A\&A, 2004, 413, 807
\bibitem[Krawczynski et al.(2002)]{krawczynski2002501model} Krawczynski, H., Coppi, P. S., \& Aharonian, F. A., MNRAS, 2002, 336, 721
\bibitem[Krawczynski et al.(2004)]{krawczynski2004orphan1959} Krawczynski, H., Hughes, S. B., Horan, D., et al., ApJ., 2004, 601, 151
\bibitem[Krennrich et al.(2002)]{krennrich2002whipple421} Krennrich, F. Bond, I. H., Bradbury, S. M., et al., ApJL, 2002, 575, 9
\bibitem[Laurent-Muehleisen et al.(1993)]{laurentmuehleisen1993vla} Laurent-Muehleisen, S. A:, Kollgaard, R. I., Moellenbrock, G. A., \& Feigelson, E. D., AJ, 1993, 106, 875 
\bibitem[Lemoine et al.(2006)]{lemoine20063dmodel} Lemoine-Goumard, M., Degrange, B., Tluczykont, M., 2006, Astropart. Phys., 25, 195
\bibitem[Lin et al.(1996)]{lin1996egret} Lin, Y. C., Bertsch, D. L., Dingus, B. L., et al., A\&AS, 1996, 120, 499
\bibitem[Maccagni et al.(1978)]{maccagni1978misidentification} Maccagni, D., Tarenghi, M., Cooke, B. A., et al., A\&A, 1978, 62, 127
\bibitem[Madau \& Pozzetti(2000)]{madau2000galaxycounts} Madau, P., \& Pozzetti, L., MNRAS, 2000, 312, L9
\bibitem[Madejski \& Schwartz(1983)]{madejski1983ssceinstein} Madejski, G. M., \& Schwartz, D. A., ApJ., 1983, 275, 467
\bibitem[Malkov \& Drury(2001)]{malkov2001acctheory} Malkov, M. A., \& Drury, L. O'C., et al. , Rep. Prog. Phys., 2001, 64, 529
\bibitem[McHardy et al.(1981)]{mchardy1981arieldetection} McHardy, I. M., Lawrence, A., Pye, J. P., \& Pounds, K. A., MNRAS, 1981, 197, 893 
\bibitem[Molendi \& Sembay(2003)]{molendi2003epic} Molendi, S., \& Sembay, S., 2003, available online at {\em http://xmm.vilspa.esa.es/docs/documents/CAL-TN-0036-1-0.ps.gz}
\bibitem[Monet et al.(1998)]{monet1998usnoa2} Monet, D., Bird A., Canzian, B., et al., 1998, available online at {\em http://vizier.cfa.harvard.edu/vizier/VizieR/pmm/usno2.htx}
\bibitem[M{\"u}cke \& Protheroe(2001)]{muecke2001protonjet} M{\"u}cke, A., \& Protheroe, R. J., Astropart. Phys., 2001, 15, 121
\bibitem[Mushotzky(1977)]{mushotzky1977sscinagn} Mushotzky, R. F., Nature, 1977, 265, 225
\bibitem[Ong(2005)]{ong2005icrcreview} Ong, R., {\em Invited, Rapporteur, and Highlight Papers, Proc. 29th ICRC (Pune)}, 2005, vol. 10, p. 329, astro-ph/0605191
\bibitem[Perlman et al.(1996)]{perlman1996einsteinbllac} Perlman, E. S., Stocke, J. T., Schachter, J., et al., ApJS, 1996, 104, 251 
\bibitem[Perlman et al.(2005)]{perlman2005curvature} Perlman, E. S:, Madejski, G., Georganopoulos, M., et al., ApJ, 2005, 625, 727
\bibitem[Pesce et al.(1994)]{pesce1994nearbygalaxies} Pesce, J. E., Falomo, R., \& Treves, A., AJ, 1994, 107, 494
\bibitem[Pinkau(1980)]{pinkau1980ngc4151} Pinkau, K., A\&A, 1980, 87, 192
\bibitem[Pohl \& Schlickeiser(2000)]{pohl2000protonjet} Pohl, M., \& Schlickeiser, R., A\&A, 2000, 354, 395
\bibitem[Primack et al.(2001)]{primack2001ebl} Primack, J. R., Somerville, R. S., Bullock, J. S, \&  Devriendt, J. E. G., Probing Galaxy formation with high energy gamma rays, AIP Conf. Proc., 2001, 558, 463
\bibitem[Primack et al.(2005)]{primack2005ebl} Primack, J. R., Bullock, J. S, \& Somerville, R. S., Observational Gamma-ray Cosmology, AIP Conf. Proc., 2005, 745, 23
\bibitem[Remillard et al.(1989)]{remillard1989opticalidentification} Remillard, R. A., Tuhoy, I. R., Brissenden, R. J. V., et al., ApJ, 1989, 345, 140
\bibitem[Romero et al.(1999)]{romero1999microvariability} Romero, G. E., Cellone, S. A., \& Combi, J. A., ApJS, 1999, 135, 477
\bibitem[Saug{\'e} \& Henri(2004)]{sauge2004pileup501} Saug{\'e}, L., \& Henri, G., ApJ, 2004, 616, 136
\bibitem[Scarpa \& Falomo(1997)]{scarpa1997opticalspectra} Scarpa, R., \& Falomo, R., A\&A, 1997, 325, 109
\bibitem[Schlegel et al.(1998)]{schlegel1998extinction} Schlegel, D. J., Finkbeiner, D. P., Davis, M., ApJ., 1998, 500, 525
\bibitem[Schlickeiser(1985)]{schlickeiser1985pileup} Schlickeiser, R., A\&A, 1985, 143, 431
\bibitem[Sikora et al.(1994)]{sikora1994ecblazarmodel} Sikora, M., Begelmann, M. C., \& Rees, M., J., ApJ., 1994, 421, 153
\bibitem[Sikora et al.(1997)]{sikora1997ssclumi} Sikora, M., Madejski, G., Moderski, R., \& Poutanen, J., ApJ., 1997, 484, 108
\bibitem[Smith et al.(1991)]{smith1991opticalcalibrators} Smith, P. S:, Jannuzi, B. T., \& Elston, R., ApJS, 1991, 77, 67
\bibitem[Stecker et al.(2006)]{stecker2006ebl} Stecker, F. W., Malkan, M. A., \& Scully, S. T., ApJ, 2006, 648, 774
\bibitem[Stecker \& Scully(2006)]{stecker2006deltagamma} Stecker, F. W., \& Scully, S. T., ApJ, 2006, 652, L9
\bibitem[Totani et al.(2001)]{totani2001ebl} Totani, T., Yoshii, Y, Iwamuro, F., et al., ApJ, 2001, 550, L137
\bibitem[Wilms et al.(2000)]{wilms2000abundances} Wilms, J., Allen, A., \& McCray, R., ApJ., 2000, 542, 914
\bibitem[Wolter et al.(1998)]{wolter1998bepposax} Wolter, A., Comastri, A., Ghisellini, G., et al., A\&A, 1998, 335, 899 
\bibitem[Wolter et al.(2000)]{wolter20001101} Wolter, A., Tavecchio, F., Caccianiga, A., Ghisellini, G., \& Tagliaferri, G., A\&A, 2000, 357, 429 
\bibitem[Xue \& Cui(2005)]{xue2005xte} Xue, Y., \& Cui, W., ApJ, 2005, 622, 160
\bibitem[Zacharias et al.(2004)]{zacharias2004ucac2} Zacharias, N., Urban, S. E., Zacharias, M. I., et al., AJ, 2004, 127, 3043
\end{thebibliography}

\expandafter\ifx\csname natexlab\endcsname\relax\def\natexlab#1{#1}\fi

\end{document}